\documentclass[aps,prd,showpacs,twocolumn,nofootinbib]{revtex4-1}

\usepackage{graphicx}
\usepackage[colorlinks=true, linkcolor=blue, citecolor=blue, urlcolor=blue]{hyperref}

\begin{document}

\title{Detailed Comparison of Renormalization Scale-Setting Procedures based on the Principle of Maximum Conformality}

\author{Xu-Dong Huang$^{1,2}$}
\email{hxud@cqu.edu.cn}

\author{Jiang Yan$^1$}
\email{yjiang@cqu.edu.cn}

\author{Hong-Hao Ma$^3$}
\email{mahonghao@pku.edu.cn}

\author{Leonardo Di Giustino$^4$}
\email{ldigiustino@uninsubria.it}

\author{Jian-Ming Shen$^5$}
\email{shenjm@hnu.edu.cn}

\author{Xing-Gang Wu$^{1}$}
\email{wuxg@cqu.edu.cn}

\author{Stanley J. Brodsky$^{6}$}
\email{sjbth@slac.stanford.edu}

\affiliation{$^1$ Department of Physics, Chongqing Key Laboratory for Strongly Coupled Physics, Chongqing University, Chongqing 401331, P.R. China}
\affiliation{$^2$ Institute of High Energy Physics, Chinese Academy of Sciences, Beijing, 100049, P.R. China}
\affiliation{$^3$ Center for High Energy Physics, Peking University, Beijing 100871, China}
\affiliation{$^4$ Department of Science and High Technology, University of Insubria, via valleggio 11, I-22100, Como, Italy}
\affiliation{$^5$ School of Physics and Electronics, Hunan University, Changsha 410082, P.R. China}
\affiliation{$^6$ SLAC National Accelerator Laboratory, Stanford University, Stanford, California 94039, USA}

\date{\today}

\begin{abstract}

The {\it Principle of Maximum Conformality} (PMC), which generalizes the conventional Gell-Mann-Low method for scale-setting in perturbative QED to non-Abelian QCD, provides a rigorous method for achieving unambiguous scheme-independent, fixed-order predictions for physical observables consistent with the principles of the renormalization group. In addition to the original multi-scale-setting approach (PMCm), two variations of the PMC have been proposed to deal with ambiguities associated with the uncalculated higher order terms in the pQCD series, i.e. the single-scale-setting approach (PMCs) and the procedures based on ``intrinsic conformality" (PMC$_\infty$). In this paper, we will give a detailed comparison of these PMC approaches by comparing their predictions for three important quantities $R_{e^+e^-}$, $R_{\tau}$, and $\Gamma(H \to b \bar{b})$ up to four-loop pQCD corrections. The PMCs approach determines an overall effective running coupling $\alpha_s(Q)$ by the recursive use of the renormalization group equation, whose argument $Q$ represents the actual momentum flow of the process. Our numerical results show that the PMCs method, which involves a somewhat simpler analysis, can serve as a reliable substitute for the full multi-scale PMCm method, and that it leads to more precise pQCD predictions with small residual scale dependence.

\end{abstract}


\maketitle

\section{Introduction}

In Quantum Chromodynamics (QCD), scattering amplitudes and cross sections at high momentum transfer can be calculated perturbatively, due to the asymptotic freedom property of the QCD running coupling~\cite{Gross:1973id, Politzer:1973fx}. The renormalization procedure for quantum field theory has been developed to make sense of the divergent integrals which emerge in perturbation calculations. The prediction for a renormalized pQCD approximant at any finite order satisfies ``renormalization group invariance" (RGI)~\cite{Petermann:1953wpa, GellMann:1954fq, Peterman:1978tb, Callan:1970yg, Symanzik:1970rt}; thus predictions for physical observables will be independent of theoretical conventions such as the choices of renormalization scheme and renormalization scale. However, if the renormalization scale for the QCD running coupling $\alpha_s$ value is instead set arbitrarily, predictions relating observables will depend on the theorist's choice of the renormalization scheme due to the mismatching of the perturbative coefficients and the $\alpha_s$ at any order, and the RGI will be violated for fixed-order predictions~\cite{Wu:2013ei, Wu:2014iba}. There are thus renormalization scale and scheme ambiguities when one uses conventional scale-setting approaches; e.g., if one chooses the renormalization scale to eliminate large logarithms, or to minimize the contributions from the higher-order loop diagrams, or to achieve theoretical predictions in agreement with the data, and etc. It is clearly important to have a rigorous scale-setting approach for achieving renormalization scale-and-scheme invariant fixed-order predictions, for deriving precise tests of the standard model (SM) at high-energy colliders such as the Large Hadron Collider (LHC), and for increasing the sensitivity of experimental measurements to new physics.

The {\it Principle of Maximum Conformality} (PMC)~\cite{Brodsky:2011ta, Brodsky:2011ig, Brodsky:2012rj, Mojaza:2012mf, Brodsky:2013vpa} is a systematic scale-setting approach satisfying all of the requirements of RGI. The PMC provides a rigorous method for obtaining unambiguous fixed-order pQCD predictions consistent with the principles of renormalization group. The purpose of PMC is not to choose an optimal renormalization scale but to determine the effective coupling constant of the process (whose argument is called as the PMC scale) with the help of renormalization group equation (RGE). It extends the Brodsky-Lepage-Mackenzie method~\cite{Brodsky:1982gc} for scale-setting in pQCD to all orders. The resulting relations among various predictions of the physical observables, such as commensurate scale relations~\cite{Brodsky:1994eh, Huang:2020gic}, ensure the PMC predictions be independent of the choice of renormalization scheme. By using the PMC, the effective coupling constant is fixed by requiring that all the RG-involved $\{\beta_i\}$-terms are eliminated. Since the effective coupling is invariant to any choice of the renormalization scale, thus solving the conventional scale ambiguity. The value of the running coupling is determined by RGE or the $\beta$-function,
\begin{equation}
\beta(a_s)=\frac{d a_s(\mu_r)} {d\ln\mu^2_r}=-a_s^2(\mu_r) \sum_{i=0}^\infty \beta_i a^{i}_s(\mu_r), \label{rgealpha}
\end{equation}
where $a_s(\mu_r)=\alpha_s(\mu_r)/(4\pi)$ and $\mu_r$ is the renormalization scale. At present, the $\{\beta_i\}$-functions in QCD are known up to five-loop level in the $\overline{\rm MS}$-scheme~\cite{Gross:1973ju, Politzer:1974fr, Caswell:1974gg, Tarasov:1980au, Larin:1993tp, vanRitbergen:1997va, Chetyrkin:2004mf, Czakon:2004bu, Baikov:2016tgj}, e.g. $\beta_0=(11-{2}n_f/{3})$, where $n_f$ is the number of active light flavors. The PMC shifts all the RG-involved non-conformal $\{\beta_i\}$-terms into $\alpha_s$ at each order of the pQCD series. It reduces to the standard scale-setting procedure of Gell-Mann and Low~\cite{GellMann:1954fq} in the QED Abelian limit (small number of colors, $N_C \to 0$~\cite{Brodsky:1997jk}). It has also been demonstrated that the PMC prediction satisfies the self-consistency conditions of the renormalization group, such as reflectivity, symmetry, and transitivity~\cite{Brodsky:2012ms}. As an important byproduct, because the RG-involved factorially divergent renormalon terms such as $n! \beta_0^n \alpha_s^n$~\cite{Beneke:1994qe, Neubert:1994vb, Beneke:1998ui} are eliminated, the convergence of the PMC perturbative series is automatically improved.

The PMC was originally introduced as a multi-scale-setting approach (PMCm)~\cite{Brodsky:2011ta, Brodsky:2012rj, Mojaza:2012mf, Brodsky:2013vpa}, in which distinct PMC scales at each order are systematically determined in order to absorb different categories of $\{\beta_i\}$-terms into the corresponding running coupling $\alpha_s$. Since the same type of $\{\beta_i\}$-terms emerge at different orders, the PMC scales at each order can be expressed in perturbative form. The precision of the PMC scale for higher-order terms decreases at high perturbative order since fewer $\{\beta_i\}$-terms are known. The PMCm has thus two kinds of residual scale dependence due to the unknown perturbative terms~\cite{Zheng:2013uja}; i.e., the last terms of the PMC scales are unknown (\textit{first kind of residual scale dependence}), and the last terms in the pQCD approximant are not fixed since its PMC scale cannot be determined (\textit{second kind of residual scale dependence}). It is noted that two of the three PMC ambiguities identified in Ref.\cite{Chawdhry:2019uuv} correspond exactly to these two kinds of residual scale dependence. These two residual scale dependence are suppressed at high orders in $\alpha_s$ and/or from exponential suppression; however, if the pQCD convergence of the perturbative series of either the PMC scale or the pQCD approximant is weak, such residual scale dependence could be significant. A recent discussion of the residual scale dependence of PMCm predictions can be found in the review~\cite{Wu:2019mky}.

The PMC single-scale-setting approach (PMCs)~\cite{Shen:2017pdu} has been suggested in order to suppress the residual scale dependence and to make the scale-setting procedures much simpler than PMCm. The PMCs procedure determines a single overall effective $\alpha_s$ with the help of RGE; the resulting PMC renormalization scale represents the overall effective momentum flow of the process. The PMCs is equivalent to PMCm in the sense of perturbative theory and the PMCs prediction is also free of renormalization scale-and-scheme ambiguities up to any fixed order~\cite{Wu:2018cmb}. By using the PMCs, the \textit{first kind of residual scale dependence} will be suppressed due to $\alpha_s$-power suppression and the exponential suppression; the overall PMC scale has the same precision at all orders, and the \textit{second kind of residual scale dependence} is exactly removed. Moreover, due to the independence on the renormalization scheme and scale, the resulting conformal series not only provides precise pQCD predictions; it also can reliably estimate the contributions from the unknown higher-order terms. The PMCs has been successfully applied to several high-energy processes~\cite{Du:2018dma, Yu:2018hgw, Yu:2019mce, Huang:2020rtx, Yu:2020tri, Huang:2020skl, Huang:2021kzc}, where the contributions from unknown higher-order terms have been estimated by using the $\rm Pad\acute{e}$ approximation approach~\cite{Basdevant:1972fe, Samuel:1992qg, Samuel:1995jc}. The PMCs also provides a self-consistency way to achieve precise $\alpha_s$ running behavior in both the perturbative and nonperturbative domains with a smooth transition from small to large scales~\cite{Deur:2017cvd, Yu:2021yvw}.

Recently, a new multi-scale PMC approach based on the principle of intrinsic conformality (iCF)~\cite{DiGiustino:2020fbk} (PMC$_\infty$) has been suggested. The PMC$_\infty$ sets the scales by requiring that all the scale-dependent $\{\beta_i\}$-terms at each order vanish separately; this provides the intrinsic conformal coefficients in an order-by-order manner. Using the PMC$_\infty$, the PMC scales at each orders are no longer evaluated as a perturbative series, thus avoiding the \textit{first kind of residual scale dependence}. The \textit{second kind of residual scale dependence} is still present, since the scale of the highest perturbative terms cannot be determined. At present, the PMC$_\infty$ has been used to predict the thrust, the wide jet broadening, the total jet broadening and the $C$-parameter distributions in $e^+ e^-$ annihilation~\cite{DiGiustino:2020fbk, DiGiustino:2021nep, Wang:2021tak}; the results are in good agreement with the data. An alternative multi-scale PMC approach (PMCa)~\cite{Chawdhry:2019uuv} has also been suggested in the literature, which as shall be shown below, it is in fact equivalent to PMC$_\infty$.

The remaining parts of this paper are organized as follows. For self-consistency, we give a mini-review of various PMC approaches such as the PMCm, the PMCs, the PMC$_\infty$, and the PMCa in Sec.II. In Sec.III, we give the numerical predictions and discussions for three observables, $R_{e^+ e^-}$, $R_{\tau}$, and $\Gamma(H\to b\bar{b})$, using each of these approaches. Section IV is reserved for a summary.

\section{A mini-review of the PMC scale-setting approaches}

In general, the pQCD prediction for a physical observable ($\rho(Q)$) can be written as the following expansion
\begin{equation}
\rho(Q)= \sum^{n}_{i=1} \left(\sum^{i-1}_{j=0} c_{i,j}(\mu_r, Q) n_f^{j}\right) a_s^{p+i-1}(\mu_r), \label{nf}
\end{equation}
where $Q$ represents a physical scale of the measured observable, $n_f$ is the number of active quark flavors, and $p$ is the power of $\alpha_s$ associated with the tree-level terms.

\subsection{The standard PMC multi-scale approach (PMCm)}
\label{pmcmsub}

The PMCm determines the precise value of $\alpha_s$ via a systematic and process-independent way, by eliminating the $\{\beta_i\}$-terms which govern the $\alpha_s$-running behavior, and by the recursive use of the RGE. To apply the PMCm procedures for the pQCD approximant (\ref{nf}), the $\{n_f\}$-series at each order must be transformed to the $\{\beta_i\}$-series~\footnote{As a subtle point, it is not simply to transform all $n_f$-terms into $\{\beta_i\}$-functions and only those that rightly govern the running behavior of the coupling constant should be transformed~\cite{Wang:2013bla}.}. The QCD degeneracy relations~\cite{Bi:2015wea} ensure the realizability of such transformation. For example, Eq.(\ref{nf}) can be rewritten as~\cite{Mojaza:2012mf, Brodsky:2013vpa}
\begin{eqnarray}
\rho(Q)&=&r_{1,0}a_s(\mu_r) + (r_{2,0}+\beta_{0}r_{2,1})a_{s}^{2}(\mu_r)\nonumber\\
&&+(r_{3,0}+\beta_{1}r_{2,1}+ 2\beta_{0}r_{3,1}+ \beta_{0}^{2}r_{3,2})a_{s}^{3}(\mu_r)\nonumber\\
&& +(r_{4,0}+\beta_{2}r_{2,1}+ 2\beta_{1}r_{3,1} + \frac{5}{2}\beta_{1}\beta_{0}r_{3,2} \nonumber\\
&& +3\beta_{0}r_{4,1}+3\beta_{0}^{2}r_{4,2}+\beta_{0}^{3}r_{4,3}) a_{s}^{4}(\mu_r)+\cdots, ~\label{rij1}
\end{eqnarray}
where $r_{i,j}$ can be derived from $c_{i,j}$. $r_{i,0}$ are conformal coefficients, and $r_{i,j}$ $(j\neq0)$ are nonconformal ones. For definiteness and without loss of generality, we have set $p=1$ and $n=4$ in order to illustrate the PMC procedure. Different types of $\{\beta_i\}$-terms can be absorbed into $\alpha_s$ via an order-by-order manner by using the RGE, which leads to distinct PMC scales at each order:
\begin{eqnarray}
a^i_s(Q_i) &\leftarrow& a^i_s(\mu_r)\bigg\{1 + i \beta_0 \frac{r_{i+1,1}}{r_{i,0}} a_s(\mu_r) \nonumber \\
&&+i\left(\beta_1 \frac{r_{i+1,1}}{r_{i,0}} +\frac{i+1}{2} \beta_0^2 \frac{r_{i+2,2}}{r_{i,0}} \right) a^{2}_s(\mu_r)\nonumber \\
&&+i\bigg(\beta_2 \frac{r_{i+1,1}}{r_{i,0}}+\frac{2i+3}{2}\beta_0\beta_1 \frac{r_{i+2,2}}{r_{i,0}}\nonumber \\
&&+\frac{(i+1)(i+2)}{3!}\beta_0^3 \frac{r_{i+3,3}}{r_{i,0}}\bigg) a^{3}_s(\mu_r)+\cdots \bigg\}. \label{scaledis1}
\end{eqnarray}
Such skeleton-like expansion~\cite{Lu:1991yu}, inversely, indicates that the PMC scales $Q_i$ should be of perturbative nature. The coefficients $r_{i,j}$ are generally function of the renormalization scale $\mu_r$, which can be redefined as
\begin{eqnarray}
r_{i,j}=\sum^j_{k=0}C^k_j{\hat r}_{i-k,j-k}{\rm ln}^k(\mu_r^2/Q^2),~\label{rijrelation}
\end{eqnarray}
where the reduced coefficients ${\hat r}_{i,j}=r_{i,j}|_{\mu_r=Q}$ (specially, we have ${\hat r}_{i,0}=r_{i,0}$), and the combination coefficients $C^k_j=j!/(k!(j-k)!)$. We then obtain
\begin{eqnarray}
\ln\frac{Q^2}{Q_1^2}&=& \frac{{\hat r}_{2,1}}{{\hat r}_{1,0}}+\beta_0\left(\frac{{\hat r}_{1,0}{\hat r}_{3,2}-{\hat r}_{2,1}^2}{{\hat r}_{1,0}^2}\right)a_s(\mu_r)\nonumber \\
&&+\bigg[\beta_1\left(\frac{3{\hat r}_{3,2}}{2r_{1,0}}-\frac{3{\hat r}_{2,1}^2}{2{\hat r}_{1,0}^2}\right)+\beta_0^2\bigg(\frac{{\hat r}_{4,3}}{{\hat r}_{1,0}}-\frac{2{\hat r}_{3,2}{\hat r}_{2,1}}{{\hat r}_{1,0}^2}\nonumber \\
&&+\frac{{\hat r}_{2,1}^3}{{\hat r}_{1,0}^3}+\frac{{\hat r}_{3,2}\ln\frac{\mu_r^2}{Q^2}}{{\hat r}_{1,0}}-\frac{{\hat r}_{2,1}^2\ln\frac{\mu_r^2}{Q^2}}{{\hat r}_{1,0}^2}\bigg)\bigg]a^2_s(\mu_r)+\cdots, \label{Q1} \\
\ln\frac{Q^2}{Q_2^2}&=& \frac{{\hat r}_{3,1}}{{\hat r}_{2,0}}+3\beta_0\frac{{\hat r}_{2,0}{\hat r}_{4,2}-{\hat r}_{3,1}^2}{2{\hat r}_{2,0}^2}a_s(\mu_r)+\cdots, \label{Q2}\\
\ln\frac{Q^2}{Q_3^2}&=& \frac{{\hat r}_{4,1}}{{\hat r}_{3,0}}+\cdots. \label{Q3}
\end{eqnarray}
These PMC scales are of a perturbative nature; one thus needs to know more higher-loop contributions in order to achieve a more precise prediction. The PMC resumes all the known same type of $\{\beta_i\}$-terms to form precise PMC scales at each order. The precision of the PMC scale for the high-order terms decreases at higher-and-higher orders due to less known $\{\beta_i\}$-terms in those higher-order terms. For example, $Q_1$ is determined up to next-to-next-to-leading log (N$^2$LL)-accuracy, $Q_2$ is determined up to NLL-accuracy and $Q_3$ is determined at the LL-accuracy. Thus the PMC scales at higher-orders are of less accuracy since more of its perturbative terms are unknown. This perturbative property of the PMC scale causes the \textit{first kind of residual scale dependence}.

After fixing the magnitude of $a_s(Q_i)$, we achieve a conformal series
\begin{eqnarray}
\rho(Q) &=& \sum_{i=1}^{4}{\hat r}_{i,0} a^i_s(Q_i) +\cdots. \label{PMCmseries}
\end{eqnarray}
The PMC scale for the highest-order term, e.g. $Q_4$ for the present case, is unfixed, since there is no $\{\beta_i\}$-terms to determine its magnitude. This makes the last perturbative term unfixed, and causes the \textit{second kind of residual scale dependence}. The PMCm usually matches $Q_4$ to the latest determined scale $Q_{3}$, which ensures the scheme independence of the prediction due to the commensurate scale relations among the predictions using different renormalization schemes~\cite{Brodsky:1994eh, Huang:2020gic}. Practically, other choices of setting $Q_4$ as a combination of various determined scales $Q_i$, such as $2 Q_3$, $(1/2) Q_3$, $(Q_1 +Q_2 +Q_3)/3$, and etc., are equally valid and will be explored in section III.

The pQCD series (\ref{PMCmseries}) is renormalization scheme-and-scale independent, and it becomes convergent due to the elimination of divergent renormalon terms. Thus a precise pQCD prediction can be achieved by applying the PMCm. Residual scale dependence arise due to perturbative nature of either the pQCD approximant or the PMC scale, which are distinct from the conventional arbitrary scale dependence. In practice, we have found that those two residual scale dependence are quite small even at low orders. This is due to the generally faster pQCD convergence after applying the PMCm~\footnote{There are cases which accidently have large cancellations among the conformal and non-conformal coefficients for a specific choice of scale under conventional scale-setting approach, and the elimination of the renormalon terms alone may lead to a even weaker convergence than conventional series. This however does not mean that conventional series is better than the PMC one; since the PMC series is scale-invariant and represents the intrinsic nature of the pQCD series, while the conventional series is largely scale dependent and the large cancellation for a specific scale shall usually disappear by choosing another scale.}. Some examples can be found in Ref.\cite{Wu:2015rga}.

\subsection{The PMC single-scale approach (PMCs)}

In some cases, the perturbative convergence may be weak in the pQCD approximant or in the series for the PMC scale, leading to a comparatively larger residual scale dependence. The PMCs procedure has been suggested to suppress this residual scale dependence by directly fixing a single effective scale and value for $\alpha_s$. Following the standard procedure of the PMCs~\cite{Shen:2017pdu}, the pQCD approximant (\ref{rij1}) changes to the following conformal series,
\begin{eqnarray}
\rho(Q)&=& \sum_{i=1}^{4} {\hat r}_{i,0}a^i_s(Q_*) + \cdots. \label{conformal}
\end{eqnarray}
The PMC scale $Q_*$ can be determined by requiring all the nonconformal terms be vanished, which can be fixed up to N$^2$LL-accuracy for $p=1$ and $n=4$. More explicitly, by requiring all the RG-involved $\{\beta_i\}$-terms to vanish, $\ln Q^2_* / Q^2$ becomes a perturbative expansion over $a_s(Q_*)$,
\begin{eqnarray}
\ln\frac{Q^2_*}{Q^2}=S_0+S_1 a_s(Q_*)+S_2 a^2_s(Q_*)+ \cdots,
\label{qstarS0}
\end{eqnarray}
whose first three expansion coefficients $S_i~(i=0, 1, 2)$ are
\begin{eqnarray}
S_0&=&-\frac{{\hat r}_{2,1}}{{\hat r}_{1,0}}, \\
S_1&=&\frac{ \beta _0 ({\hat r}_{2,1}^2-{\hat r}_{1,0} {\hat r}_{3,2})}{{\hat r}_{1,0}^2}+\frac{2 ({\hat r}_{2,0} {\hat r}_{2,1}-{\hat r}_{1,0} {\hat r}_{3,1})}{{\hat r}_{1,0}^2},
\end{eqnarray}
and
\begin{eqnarray}
S_2&&=\frac{3 \beta _1 ({\hat r}_{2,1}^2-{\hat r}_{1,0} {\hat r}_{3,2})}{2 {\hat r}_{1,0}^2}\nonumber\\
&&+\frac{4({\hat r}_{1,0} {\hat r}_{2,0} {\hat r}_{3,1}-{\hat r}_{2,0}^2 {\hat r}_{2,1})+3({\hat r}_{1,0} {\hat r}_{2,1} {\hat r}_{3,0}-{\hat r}_{1,0}^2 {\hat r}_{4,1})}{ {\hat r}_{1,0}^3} \nonumber \\
&&+\frac{ 3\beta_0 (2{\hat r}_{2,1} {\hat r}_{3,1} {\hat r}_{1,0}- {\hat r}_{4,2} {\hat r}_{1,0}^2)}{ {\hat r}_{1,0}^3}\nonumber\\
&&+\frac{ \beta _0 (2 {\hat r}_{2,0} {\hat r}_{3,2} {\hat r}_{1,0}-5 {\hat r}_{2,0} {\hat r}_{2,1}^2)}{ {\hat r}_{1,0}^3}\nonumber\\
&&+\frac{ \beta _0^2 (3 {\hat r}_{1,0} {\hat r}_{3,2} {\hat r}_{2,1}-2 {\hat r}_{2,1}^3- {\hat r}_{1,0}^2 {\hat r}_{4,3})}{ {\hat r}_{1,0}^3}.
\end{eqnarray}
Then, one can derive the exact value of $Q_*$ by solving Eq.(\ref{qstarS0}) numerically.

Eq.(\ref{qstarS0}) shows that the exponential form of the PMC scale, $\ln Q^2_*/Q^2$, is a power series in $\alpha_s$, which resums all the known $\{\beta_i\}$-terms via the RGE, and is independent of renormalization scale $\mu_r$ at any fixed order. It represents the correct momentum flow of the process and determines an overall effective value of $\alpha_s$. Together with the $\mu_r$-independent conformal coefficients, the resulting pQCD series is exactly scheme-and-scale independent~\cite{Wu:2018cmb}. By using the PMCs approach, the PMC scale-setting procedure is greatly simplified -- there is no \textit{second kind of residual scale dependence}. And for a general $a_s^n$-order pQCD series similar to (\ref{rij1}), the \textit{first kind of residual scale dependence} for the PMCm series (\ref{PMCmseries}) takes the form, $\propto \left(\sum\limits_{i=1}^{n} \hat{r}_{i,0}\right){\cal O}(a_s^{n+1})$, which changes to the one with a smaller magnitude, $\propto \left(\sum\limits_{i=1}^{n} \hat{r}_{i,0}a_s^{i-1}\right){\cal O}(a_s^{n+1})$, for the PMCs series (\ref{conformal}).

\subsection{The PMC multi-scale approach using the intrinsic conformality (PMC$_\infty$)}
\label{subsec-c}

In Ref.\cite{DiGiustino:2020fbk}, it has been pointed out that the renormalizable SU(N)/U(1) gauge theories have the property of intrinsic conformality (iCF) which ensures the scale invariance of the pQCD series at each perturbative order. The iCF thus yields a particular structure of the perturbative corrections, and the pQCD approximant (\ref{nf}) can be rewritten as
\begin{eqnarray}
\rho(Q)&=& \sum_{i=1}^{4} \mathcal{C}_{i}(\mu_r, Q)a^i_s(\mu_r)+ \cdots, \label{nfseries} \\
&=& \mathcal{C}_{1,\rm{IC}} a_s(\mu_{\rm I}) + \mathcal{C}_{2,\rm{IC}} a^2_s(\mu_{\rm II}) \nonumber\\
& & + \mathcal{C}_{3,\rm{IC}} a^3_s(\mu_{\rm III})+ \mathcal{C}_{4,\rm{IC}} a^4_s(\mu_{\rm IV})+\cdots, \label{PMCinfseries}
\end{eqnarray}
where $\mathcal{C}_{i,\rm{IC}}$ are the intrinsic conformal (IC) coefficients free of scale dependence, which can be derived from the known coefficients $\mathcal{C}_{i}(\mu_r, Q)$ by using the property of intrinsic conformality. The second equation is the perturbative series satisfies the iCF, being scale invariant at each order, which can be transformed back to the first equation by using the scale displacement relation which can be simply derived from the RGE (\ref{rgealpha}). Then we can obtain the following relations:
\begin{eqnarray}
\mathcal{C}_{1}(\mu_r)&=&\mathcal{C}_{1,\rm{IC}},\\
\mathcal{C}_{2}(\mu_r)&=&\mathcal{C}_{2,\rm{IC}}+\ln \frac{\mu_r^2}{\mu_{\rm I}^2}\beta_0\mathcal{C}_{1,\rm{IC}},\label{Amu0}\\
\mathcal{C}_{3}(\mu_r)&=&\mathcal{C}_{3,\rm{IC}}+2\ln\frac{\mu_r^2}{\mu_{\rm II}^2}\beta_0\mathcal{C}_{2,\rm{IC}}\nonumber \\ &&+\left( \ln\frac{\mu_r^2}{\mu_{\rm I}^2}\beta_0^2+\beta_1\right)\ln\frac{\mu_r^2}{\mu_{\rm I}^2}\mathcal{C}_{1,\rm{IC}},\\
\mathcal{C}_{4}(\mu_r)&=&\mathcal{C}_{4,\rm{IC}}+\frac{5}{2}\bigg(\ln\frac{\mu_r^2}{\mu_{\rm I}^2}\bigg)^2\beta_0\beta_1\mathcal{C}_{1,\rm{IC}}\nonumber \\&&+\ln\frac{\mu_r^2}{\mu_{\rm I}^2}\beta_2\mathcal{C}_{1,\rm{IC}}+2\ln\frac{\mu_r^2}{\mu_{\rm II}^2}\beta_1\mathcal{C}_{2,\rm{IC}}\nonumber \\&&+3\ln\frac{\mu_r^2}{\mu_{\rm III}^2}\beta_0\mathcal{C}_{3,\rm{IC}}+3\bigg(\ln\frac{\mu_r^2}{\mu_{\rm II}^2}\bigg)^2\beta_0^2\mathcal{C}_{2, \rm{IC}}\nonumber \\ &&+\bigg(\ln\frac{\mu_r^2}{\mu_{\rm I}^2}\bigg)^3\beta_0^3\mathcal{C}_{1,\rm{IC}}. \label{Dmu0}
\end{eqnarray}
The iCF ensures that each subset of the perturbative series to be scale-invariant, and inversely, that each subset corresponds to the same IC coefficient. We then need to fix a different scale for each order. The PMC$_\infty$ scale-setting procedures can be done via an order-by-order manner: the first IC coefficient $\mathcal{C}_{1,\rm{IC}}$ can be easily determined, which directly equals to $\mathcal{C}_{1}$; the second IC coefficient $\mathcal{C}_{2,\rm{IC}}$ can be fixed by setting $n_f\equiv {33}/{2}$~\footnote{As a subtle point, the condition of $n_f\equiv {33}/{2}$ should be applied to the renormalization group involved $n_f$-terms only~\cite{Wu:2013ei}, which ensures the equivalence of the conformal coefficients among PMC$_\infty$ and PMCm predictions.} in $\mathcal{C}_{2}$, which leads to $\beta_0=0$, and then all scale-dependent terms are eliminated at this order; i.e. $\mathcal{C}_{2,\rm{IC}}=\mathcal{C}_{2}|_{n_f={33}/{2}}$, etc. More explicitly, we have
\begin{eqnarray}
\mathcal{C}_{1,\rm{IC}}&=&\mathcal{C}_{1}, \label{Aconf}\\
\mathcal{C}_{2,\rm{IC}}&=&\mathcal{C}_{2}|_{n_f=\frac{33}{2}},\\
\mathcal{C}_{3,\rm{IC}}&=&\mathcal{C}_{3}|_{n_f=\frac{33}{2}}-\bar{\beta}_1\ln\frac{\mu_r^2}{\mu_{\rm I}^2}{\bigg |}_{n_f=\frac{33}{2}}\mathcal{C}_{1,\rm{IC}},\\
\mathcal{C}_{4,\rm{IC}}&=&\mathcal{C}_{4}|_{n_f=\frac{33}{2}}-\bar{\beta}_2\ln\frac{\mu_r^2}{\mu_{\rm I}^2}{\bigg |}_{n_f=\frac{33}{2}}\mathcal{C}_{1,\rm{IC}}\nonumber\\
&&-2\bar{\beta}_1\ln\frac{\mu_r^2}{\mu_{\rm II}^2}{\bigg |}_{n_f=\frac{33}{2}}\mathcal{C}_{2,\rm{IC}}.
\end{eqnarray}
where the $\bar{\beta}_1=\beta_1|_{n_f=\frac{33}{2}}$, and $\bar{\beta}_2=\beta_2|_{n_f=\frac{33}{2}}$. The first three PMC$_\infty$ scales take the following forms:
\begin{eqnarray}
\ln\frac{\mu_r^2}{\mu_{\rm I}^2}&=&\frac{\mathcal{C}_{2}-\mathcal{C}_{2,\rm{IC}}}{\beta_0\mathcal{C}_{1,\rm{IC}}},\\
\ln\frac{\mu_r^2}{\mu_{\rm II}^2}&=&\bigg\{\mathcal{C}_{3}-\mathcal{C}_{3,\rm{IC}}-\left[ \ln\frac{\mu_r^2}{\mu_{\rm I}^2}\beta_0^2+\beta_1\right]\times \nonumber \\
&&\ln\frac{\mu_r^2}{\mu_{\rm I}^2}\mathcal{C}_{1,\rm{IC}}\bigg\}/(2\beta_0\mathcal{C}_{2,\rm{IC}}),\\
\ln\frac{\mu_r^2}{\mu_{\rm III}^2}&=&\bigg[\mathcal{C}_{4}-\mathcal{C}_{4,\rm{IC}}-3\bigg(\ln\frac{\mu_r^2}{\mu_{\rm II}^2}\bigg)^2\beta_0^2\mathcal{C}_{2, \rm{IC}}\nonumber \\&&-\bigg(\ln\frac{\mu_r^2}{\mu_{\rm I}^2}\bigg)^3\beta_0^3\mathcal{C}_{1,\rm{IC}}-2\ln\frac{\mu_r^2}{\mu_{\rm II}^2}\beta_1\mathcal{C}_{2,\rm{IC}}\nonumber \\&&-\frac{5}{2}\bigg(\ln\frac{\mu_r^2}{\mu_{\rm I}^2}\bigg)^2\beta_0\beta_1\mathcal{C}_{1,\rm{IC}}\nonumber \\&&-\ln\frac{\mu_r^2}{\mu_{\rm I}^2}\beta_2\mathcal{C}_{1,\rm{IC}}\bigg]/(3\beta_0\mathcal{C}_{3,\rm{IC}}).\label{uIII}
\end{eqnarray}
Further, by using the relations between $\mathcal{C}_{i}$ and $r_{i,j}$ ($\hat{r}_{i,j}$) defined in Eqs.(\ref{rij1}, \ref{rijrelation}), we obtain
\begin{eqnarray}
\mathcal{C}_{1,\rm{IC}}&=&{\hat r}_{1,0}, \\
\mathcal{C}_{2,\rm{IC}}&=&{\hat r}_{2,0}, \\
\mathcal{C}_{3,\rm{IC}}&=&{\hat r}_{3,0}, \\
\mathcal{C}_{4,\rm{IC}}&=&{\hat r}_{4,0}.
\end{eqnarray}
and
\begin{eqnarray}
\ln\frac{Q^2}{\mu_{\rm I}^2}&=& \frac{{\hat r}_{2,1}}{{\hat r}_{1,0}}, \label{Infq1} \\
\ln\frac{Q^2}{\mu_{\rm II}^2}&=& \frac{{\hat r}_{3,1}}{{\hat r}_{2,0}}+\beta_0\left(\frac{{\hat r}_{3,2}}{2{\hat r}_{2,0}}-\frac{{\hat r}_{2,1}^2}{2{\hat r}_{1,0}{\hat r}_{2,0}}\right), \label{Infq2} \\
\ln\frac{Q^2}{\mu_{\rm III}^2}&=& \frac{{\hat r}_{4,1}}{{\hat r}_{3,0}}+\beta_0\left(\frac{{\hat r}_{4,2}}{{\hat r}_{3,0}}-\frac{{\hat r}_{3,1}^2}{{\hat r}_{2,0} {\hat r}_{3,0}}\right)\nonumber \\
&+&\beta_0^2\left(\frac{{\hat r}_{4,3}}{3{\hat r}_{3,0}}-\frac{{\hat r}_{2,1}^3}{3{\hat r}_{1,0}^2 {\hat r}_{3,0}}+\frac{{\hat r}_{3,1}{\hat r}_{2,1}^2}{{\hat r}_{1,0} {\hat r}_{2,0} {\hat r}_{3,0}}-\frac{{\hat r}_{3,1} {\hat r}_{3,2}}{{\hat r}_{2,0} {\hat r}_{3,0}}\right) \nonumber \\
&+&\beta_0^3\left(\frac{{\hat r}_{3,2}{\hat r}_{2,1}^2}{2{\hat r}_{1,0} {\hat r}_{2,0} {\hat r}_{3,0}}-\frac{{\hat r}_{3,2}^2}{4{\hat r}_{2,0} {\hat r}_{3,0}}-\frac{{\hat r}_{2,1}^4}{4{\hat r}_{1,0}^2 {\hat r}_{2,0} {\hat r}_{3,0}}\right)\nonumber \\
&+&\beta_1\left(\frac{{\hat r}_{3,2}}{2{\hat r}_{3,0}}-\frac{{\hat r}_{2,1}^2}{2{\hat r}_{1,0} {\hat r}_{3,0}}\right). \label{Infq3}
\end{eqnarray}

It is noted that the IC coefficients ($\mathcal{C}_{i=(1,\cdots,4),\rm{IC}}$) are exactly equal to the PMCm and the PMCs conformal coefficients (${\hat r}_{i=(1,\cdots,4),0}$). This agreement can be extended up to all orders, since the PMC$_\infty$ approach, in effect, also eliminates the renormalization group involved non-conformal $\{\beta_i\}$-terms. And in distinction to the PMCm and PMCs, the PMC$_\infty$ scales $\mu_{\rm I}$, $\mu_{\rm II}$, $\mu_{\rm III}$ do not have a perturbative nature; thus there is no \textit{first kind of residual scale dependence} for the PMC$_\infty$ predictions. This property is evident, since the PMC$_\infty$ effective scales ensure the scale-independence of each loop term; in contrast, the effective scales of PMCm are chosen to fix the correct magnitude of the effective $\alpha_s$ (and hence the effective scale) of each loop terms by using the RGE. As is the case of PMCm, the PMC$_\infty$ prediction does not determine the scale of the last term; e.g. we need to know detailed information for ${\cal C}_5$ to set the scale $\mu_{\rm IV}$, and thus it still has the \textit{second kind of residual scale dependence}.

\subsection{An alternative method for PMC multi-scale approach (PMCa)}

As shown in Sec.II.A, in the PMCm approach, the resulting PMC scales are of perturbative nature. Ref.\cite{Chawdhry:2019uuv} suggests an alternative multi-scale-setting procedure (PMCa), whose effective scales ($q_k$), similar to the case of PMC$_\infty$, are not of the perturbative form, and thus also avoid the \textit{first kind of residual scale dependence}. More explicitly, the PMCa starts from the conformal series (\ref{PMCmseries}), i.e.
\begin{eqnarray}
\rho(Q) &=& \sum_{i=1}^{4} \hat{r}_{i,0} a^i_s(q_i) + \cdots,~\label{rij3}
\end{eqnarray}
where the conformal coefficients are the same as those of the PMCm approach, but using the effective scales $q_i$ to replace the PMCm ones $Q_i$. Using the scale displacement relation derived from the RGE (\ref{rgealpha}), one obtains
\begin{eqnarray}
\rho(Q)&=& \hat{r}_{1,0}a_s(\mu_r) + \left(\hat{r}_{2,0}+\beta_{0} \hat{r}_{1,0}\ln\frac{\mu_r^2}{q_1^2}\right)a_{s}^{2}(\mu_r) \nonumber \\ &&+\bigg[ \hat{r}_{3,0}+\beta_{1} \hat{r}_{1,0}\ln\frac{\mu_r^2}{q_1^2}+ 2\beta_{0} \hat{r}_{2,0}\ln\frac{\mu_r^2}{q_2^2}\nonumber \\
&&+\beta_{0}^{2} \hat{r}_{1,0}\left(\ln\frac{\mu_r^2}{q_1^2}\right)^2\bigg]a_{s}^{3}(\mu_r) +\bigg[\hat{r}_{4,0} \nonumber \\
&&+\beta_{2} \hat{r}_{1,0}\ln\frac{\mu_r^2}{q_1^2}+ 2\beta_{1} \hat{r}_{2,0}\ln\frac{\mu_r^2}{q_2^2} \nonumber \\
&&+\frac{5}{2}\beta_{1}\beta_{0} \hat{r}_{1,0} \left(\ln\frac{\mu_r^2}{q_1^2} \right)^2 +3\beta_{0} \hat{r}_{3,0}\ln\frac{\mu_r^2}{q_3^2}\nonumber \\&&+3\beta_{0}^{2} \hat{r}_{2,0}\left(\ln\frac{\mu_r^2}{q_2^2}\right)^2+\beta_{0}^{3} \hat{r}_{1,0}\left(\ln\frac{\mu_r^2}{q_1^2}\right)^3\bigg] a_{s}^{4}(\mu_r)\nonumber \\&&+\cdots ~\label{rij2}
\end{eqnarray}
The effective scales $q_i$ are then fixed by equating the perturbative coefficients of Eq.(\ref{rij1}) and Eq.(\ref{rij2}) to be exactly the same, which ensures the resultant pQCD series to be scale independent at each order. The derived first three ones are
\begin{eqnarray}
\ln\frac{Q^2}{q_1^2}&&= \frac{{\hat r}_{2,1}}{{\hat r}_{1,0}}, ~\label{q1}\\
\ln\frac{Q^2}{q_2^2}&&= \frac{{\hat r}_{3,1}}{{\hat r}_{2,0}}+\beta_0\left(\frac{{\hat r}_{3,2}}{2{\hat r}_{2,0}}-\frac{{\hat r}_{2,1}^2}{2{\hat r}_{1,0}{\hat r}_{2,0}}\right),~\label{q2}\\
\ln\frac{Q^2}{q_3^2}&&= \frac{{\hat r}_{4,1}}{{\hat r}_{3,0}}+\beta_0\left(\frac{{\hat r}_{4,2}}{{\hat r}_{3,0}}-\frac{{\hat r}_{3,1}^2}{{\hat r}_{2,0} {\hat r}_{3,0}}\right)\nonumber \\
&&+\beta_0^2\left(\frac{{\hat r}_{4,3}}{3{\hat r}_{3,0}}-\frac{{\hat r}_{2,1}^3}{3{\hat r}_{1,0}^2 {\hat r}_{3,0}}+\frac{{\hat r}_{3,1}{\hat r}_{2,1}^2}{{\hat r}_{1,0} {\hat r}_{2,0} {\hat r}_{3,0}}-\frac{{\hat r}_{3,1} {\hat r}_{3,2}}{{\hat r}_{2,0} {\hat r}_{3,0}}\right)\nonumber \\
&&+\beta_0^3\left(\frac{{\hat r}_{3,2}{\hat r}_{2,1}^2}{2{\hat r}_{1,0} {\hat r}_{2,0} {\hat r}_{3,0}}-\frac{{\hat r}_{3,2}^2}{4{\hat r}_{2,0} {\hat r}_{3,0}}-\frac{{\hat r}_{2,1}^4}{4{\hat r}_{1,0}^2 {\hat r}_{2,0} {\hat r}_{3,0}}\right)\nonumber \\
&&+\beta_1\left(\frac{{\hat r}_{3,2}}{2{\hat r}_{3,0}}-\frac{{\hat r}_{2,1}^2}{2{\hat r}_{1,0} {\hat r}_{3,0}}\right).~\label{q3}
\end{eqnarray}

Comparing Eqs.(\ref{Infq1}, \ref{Infq2}, \ref{Infq3}) with Eqs.(\ref{q1}, \ref{q2}, \ref{q3}), it is found that the first three effective scales of the PMCa and PMC$_\infty$ approaches are exactly the same. The agreement of scales for both approaches can be naturally extended to all orders. Thus the pQCD series (\ref{PMCinfseries}) and (\ref{rij3}) are identical, demonstrating that the PMCa and the PMC$_\infty$ approaches overlap, even though the detailed procedures for fixing the effective scales are different.

\section{Numerical Results}

In this section, we will compare the predictions for three important observables: $R_{e^+e^-}$, $R_{\tau}$, and $\Gamma(H \to b \bar{b})$, which have been calculated up to four-loop QCD corrections using alternative PMC scale-setting procedures. Numerical results for the conventional, PMCm, PMCs and PMC$_\infty$ approaches will be presented. Because the PMCa and the PMC$_\infty$ approaches are equivalent, we will only give the PMC$_\infty$ predictions. For self-consistency, the same loop $\alpha_s$-running behavior will be adopted for calculating the same loop perturbative series. The QCD asymptotic scale ($\Lambda_{\rm QCD}$) is then fixed by using $\alpha_s(M_Z)=0.1179$~\cite{Workman:2022zbs}, and for a four-loop prediction, we obtain $\Lambda_{\rm{QCD}}^{n_f=4}=291.7$ MeV and $\Lambda_{\rm{QCD}}^{n_f=5}=207.2$ MeV in conventional $\overline{\rm MS}$ renormalization scheme.

\subsection{The pQCD predictions for $R_{e^+e^-}$, $R_{\tau}$, and $\Gamma(H \to b \bar{b})$}

The annihilation of electron and positron into hadrons provides one of the most important platforms for determining the running behavior of the QCD coupling. The $R$-ratio is defined as
\begin{eqnarray}
R_{e^+ e^-}(Q)&=&\frac{\sigma\left(e^+e^-\rightarrow {\rm hadrons} \right)}{\sigma\left(e^+e^-\rightarrow \mu^+\mu^-\right)}\nonumber\\
&=& 3\sum_q e_q^2\left[1+R(Q)\right], \label{Re+e-}
\end{eqnarray}
where $Q=\sqrt s $, corresponding to the electron-positron collision energy in the center-of-mass frame. The pQCD series for $R(Q)$, up to $(n+1)$-loop QCD corrections, is
\begin{displaymath}
R_n(Q)=\sum_{i=0}^{n} {\cal C}_{i}(Q,\mu_r) a_s^{i+1}(\mu_r).
\end{displaymath}
The perturbative coefficients ${\cal C}_{i}(Q,\mu_r)$ in the $\overline{\rm MS}$-scheme up to four-loop level have been calculated in Refs.\cite{Baikov:2008jh, Baikov:2010je, Baikov:2012zm, Baikov:2012zn}. As a reference point, at $\sqrt{s} =31.6$ GeV, we have $\frac{3}{11}R_{e^+ e^-}^{\rm Exp.}=1.0527 \pm0.0050$~\cite{Marshall:1988ri}.

Another useful ratio $R_{\tau}$ for the $\tau$-lepton decays into hadrons is defined as
\begin{eqnarray}
R_{\tau}(M_{\tau})&=&\frac{\Gamma\left(\tau\rightarrow \nu_{\tau}+{\rm hadrons} \right)}{\Gamma\left(\tau\rightarrow \nu_{\tau}+ \bar{\nu}_{l}+l\right)}\nonumber\\
&=& 3|V_{ud}|^2S_{\rm EW}\left[1+\hat{R}(M_{\tau})+\delta_{\rm EW}^{'}+\delta_2+\delta_{\rm NP}\right], \label{Rtau}
\end{eqnarray}
where $V_{ud}=0.97373\pm0.00031$~\cite{Workman:2022zbs} is Cabbibo-Kobayashi-Maskawa matrix elements, $S_{\rm EW}=1.0198\pm0.0006$ and $\delta_{\rm EW}^{'}=0.001$ for the electroweak corrections, $\delta_2=(-4.4\pm2.0)\times 10^{-4}$ for light quark mass effects, $\delta_{\rm NP}=(-4.8\pm1.7)\times 10^{-3}$ for the nonperturbative effects and $M_{\tau}= 1.777$ GeV~\cite{Davier:2005xq,ALEPH:2005qgp,Baikov:2008jh}. The pQCD series for $\hat{R}(M_{\tau})$ up to $(n+1)$-loop QCD corrections is
\begin{displaymath}
\hat{R}_{n}(M_{\tau})=\sum_{i=0}^{n} \hat{\cal C}_i(M_{\tau},\mu_r) a_s^{i+1}(\mu_r),
\end{displaymath}
where the perturbative coefficients $\hat{\cal C}_i(M_{\tau},\mu_r)$ up to four-loop QCD corrections can be derived by using the relation between $R_{\tau}(M_{\tau})$ and $R_{e^+ e^-}(\sqrt{s})$~\cite{Lam:1977cu}.

The decay width for Higgs boson decay into a bottom and anti-bottom pair $H\to b\bar{b}$ can be written as
\begin{eqnarray}
\Gamma(H\to b\bar{b})=\frac{3G_{F} M_{H} m_{b}^{2}(M_{H})} {4\sqrt{2}\pi} [1+\tilde{R}(M_{H})],
\end{eqnarray}
where the Fermi constant $G_{F}=1.16638\times10^{-5}\;\rm{GeV}^{-2}$, the Higgs mass $M_{H} =125.1$ GeV, and the $b$-quark $\overline{\rm{MS}}$-running mass is $m_b(M_H)=2.78$ GeV~\cite{Wang:2013bla}. The pQCD series for $\tilde{R}(M_{H})$ up to $(n+1)$-loop QCD corrections is
\begin{displaymath}
\tilde{R}_n(M_{H})=\sum_{i=0}^{n} \tilde{\cal C}_i(M_{H},\mu_r) a_s^{i+1}(\mu_r).
\end{displaymath}
The perturbative coefficients $\tilde{\cal C}_i(M_{H},\mu_r)$ up to four-loop QCD corrections have been calculated in Ref.\cite{Baikov:2005rw}.

In the following, we give the properties for the pQCD series of $R_{n}(Q=31.6~{\rm GeV})$, $\hat{R}_{n}(M_\tau)$, and $\tilde{R}_{n}(M_H)$ using each scale-setting approach. As for the leading-order ratios with $n=0$, we have no information to set the renormalization scale for all the scale-setting approaches; and for convenience, we directly set it to be $Q$, $M_\tau$, or $M_H$, respectively, which gives $R_0=0.04428$, $\hat{R}_0=0.0891$, and $\tilde{R}_0=0.2034$. \\

\subsection{properties using the conventional scale-setting approach}

\begin{table*}[htb]
\centering
\begin{tabular}{ccccc}
\hline
& ~~~${\cal C}_1(\mu_r)$~~~ & ~~~${\cal C}_2(\mu_r)$~~~ & ~~~${\cal C}_3(\mu_r)$~~~ & ~~~${\cal C}_4(\mu_r)$~~~ \\
 \hline
 $R(Q)$ & $4$ & $22.55^{+42.51}_{-42.51}$ & $-819.50^{+1145.54}_{-266.26}$ & $-20591^{+31244.8}_{-7812.4}$ \\
 \hline
 $\hat{R}(M_{\tau})$ & $4$ & $83.24^{+49.91}_{-41.39}$ & $1687.42^{+3054.60}_{-1588.77}$ & $32532.1^{+139210.0}_{-37674.5}$ \\
 \hline
 $\tilde{R}(M_H)$ & $22.6667$ & $466.347^{+240.907}_{-240.907}$ & $2672.49^{+13688.40}_{-8567.51}$ & $-211391^{+358424}_{-27651.9}$ \\
 \hline
\end{tabular}
\caption{The scale-dependent coefficients ${\cal C}_i(\mu_r)$ of the conventional series for $R_{n}(Q=31.6~{\rm GeV})$, $\hat{R}_{n}(M_\tau)$, and $\tilde{R}_{n}(M_H)$, respectively. The central values are for $\mu_r=Q$, $M_\tau$, or $M_H$, respectively. The errors are caused by taking $\mu_r\in [1/2 Q, 2 Q]$ for $R_{n}(Q=31.6~{\rm GeV})$, $\mu_r\in [1{\rm GeV}, 2M_{\tau}]$ for $\hat{R}_{n}(M_{\tau})$, and $\mu_r\in [1/2 M_H, 2 M_H]$ for $\tilde{R}_{n}(M_H)$, respectively.} \label{cijn}
\end{table*}

We present the perturbative coefficients ${\cal C}_i(\mu_r)$ in Table~\ref{cijn}, where the errors are caused by taking $\mu_r\in [1/2 Q, 2 Q]$ for $R(Q=31.6~{\rm GeV})$, $\mu_r\in [1{\rm GeV}, 2M_{\tau}]$ for $\hat{R}(M_{\tau})$, and $\mu_r\in [1/2 M_H, 2 M_H]$ for $\tilde{R}(M_H)$, respectively. Table~\ref{cijn} shows that those coefficients are highly scale dependent.

\begin{table*}[htb]
\centering
\begin{tabular}{ccccccc}
\hline
& $n=1$ & $n=2$ & $n=3$ & $\kappa_1$ & $\kappa_2$ & $\kappa_3$ \\
 \hline
 $R_n|_{\rm Conv.}$ & $0.04753^{+0.00044}_{-0.00138}$ & $0.04638^{+0.00012}_{-0.00070}$ & $0.04608^{+0.00015}_{-0.00009}$ & $7.3^{-2.9}_{+9.2}\%$ & $2.4^{+0.7}_{-1.4}\%$ & $0.6^{-0.0}_{+0.1}\%$ \\
 \hline
 $\hat{R}_n|_{\rm Conv.}$ & $0.1522^{+0.0482}_{-0.0295}$ & $0.1826^{+0.0360}_{-0.0268}$ & $0.1980^{+0.0170}_{-0.0194}$ & $70.8^{+2.6}_{+2.3}\%$ & $20.0^{-10.9}_{+7.0}\%$ & $8.4^{-6.8}_{+6.2}\%$ \\
 \hline
 $\tilde{R}_n|_{\rm Conv.}$ & $0.2404^{+0.0074}_{-0.0075}$ & $0.2423^{+0.0002}_{-0.0007}$ & $0.2409^{+0.0015}_{-0.0007}$ & $18.2^{-8.0}_{+6.7}\%$ & $0.8^{+1.3}_{+2.9}\%$ & $0.6^{-0.6}_{+0.0}\%$ \\
 \hline
\end{tabular}
\caption{Results for $R_{n}(Q=31.6~{\rm GeV})$, $\hat{R}_{n}(M_{\tau})$, $\tilde{R}_{n}(M_H)$ up to four-loop QCD corrections using the conventional scale-setting approach. The central values are obtained by setting $\mu_r$ as $Q$, $M_\tau$, or $M_H$, respectively. The errors are caused by taking $\mu_r\in [1/2 Q, 2 Q]$ for $R_{n}(Q=31.6~{\rm GeV})$, $\mu_r\in [1{\rm GeV}, 2M_{\tau}]$ for $\hat{R}_{n}(M_{\tau})$, and $\mu_r\in [1/2 M_H, 2 M_H]$ for $\tilde{R}_{n}(M_H)$.}
\label{conv}

\end{table*}

We present the results of $R_{n}(Q=31.6~{\rm GeV})$, $\hat{R}_{n}(M_{\tau})$, and $\tilde{R}_{n}(M_H)$ up to four-loop QCD corrections using the conventional scale-setting approach in Table~\ref{conv}, where the errors are caused by taking $\mu_r\in [1/2 Q, 2 Q]$ for $R_{3}(Q=31.6~{\rm GeV})$, $\mu_r\in [1{\rm GeV}, 2M_{\tau}]$ for $\hat{R}_{3}(M_{\tau})$, and $\mu_r\in [1/2 M_H, 2 M_H]$ for $\tilde{R}_{3}(M_H)$, respectively. For self-consistency, we adopt the $(n+1)_{\rm th}$-loop $\alpha_s$-running behavior in deriving $(n+1)_{\rm th}$-loop prediction for $R_{n}(Q=31.6~{\rm GeV})$, $\hat{R}_{n}(M_{\tau})$, and $\tilde{R}_{n}(M_H)$. We define the ratio
\begin{displaymath}
\kappa_n = \left|\frac{{\cal R}_n-{\cal R}_{n-1}}{{\cal R}_{n-1}}\right|,
\end{displaymath}
where ${\cal R}$ stands for $R$, $\hat{R}$ and $\tilde{R}$, respectively. It shows how the ``known" prediction ${\cal R}_{n-1}$ is affected by the one-order-higher terms. Table~\ref{conv} shows that generally, we have $\kappa_1 >\kappa_2 >\kappa_3$ for all those quantities, which are consistent with the perturbative nature of the series and indicates that one can obtain more precise prediction by including more loop terms. To show the perturbative nature more explicitly, we present the magnitudes of each loop terms for the four-loop approximants $R_{3}(Q=31.6~{\rm GeV})$, $\hat{R}_{3}(M_{\tau})$ and $\tilde{R}_{3}(M_H)$ in Table \ref{convorder}. Table \ref{convorder} shows the relative importance among the LO-terms, the NLO-terms, the N$^2$LO-terms and the N$^3$LO-terms for those approximants are
\begin{eqnarray}
&& 1 : +0.063^{+0.099}_{-0.100} : -0.026^{+0.034}_{-0.016}  : -0.007^{+0.001}_{+0.009}, \label{conv41} \\
&& 1 : +0.531^{-0.108}_{+0.101} : +0.275^{-0.205}_{+0.152} : +0.135^{-0.186}_{+0.159}, \label{conv42}\\
&& 1 : +0.184^{+0.071}_{-0.085} : +0.009^{+0.039}_{-0.035} : -0.007^{+0.010}_{-0.002}, \label{conv43}
\end{eqnarray}
where the central values are for $\mu_r=Q$, $\mu_r=M_\tau$ and $\mu_r=M_H$; and the errors are for $\mu_r\in[Q/2, 2Q]$, $\mu_r\in[1{\rm GeV}, 2M_\tau]$ and $\mu_r\in[M_H/2, 2M_H]$, respectively. Being consistent with Table~\ref{conv}, the scale dependence for each loop terms is large, but due to the cancellation of scale dependence among different orders, the net scale dependence is small, e.g. $\left(^{+0.3\%}_{-0.2\%}\right)$, $\left(^{+8.6\%}_{-9.8\%}\right)$ and $\left(^{+0.6\%}_{-0.3\%}\right)$ for $R_{3}(Q=31.6~{\rm GeV})$, $\hat{R}_{3}(M_{\tau})$ and $\tilde{R}_{3}(M_H)$, respectively. It is noted that due to the usual renormalon divergence and a larger $\alpha_s$ value at a smaller scale $M_\tau$, i.e. $\alpha_s(M_\tau) \sim 0.33$, the net scale dependence of the four-loop prediction $\hat{R}_{3}(M_{\tau})$ is still sizable.

\begin{widetext}
\begin{center}
\begin{table}[htb]
\begin{tabular}{cccccc}
\hline
 & ~~~~$\rm LO$~~~~ & ~~~~$\rm NLO$~~~~ & ~~~~$\rm N^2LO$~~~~ & ~~~~$\rm N^3LO$~~~~ & ~~~~$\rm Total$~~~~ \\
 \hline
 ~~$R_3|_{\rm Conv.}$~~ & $0.04473^{-0.00499}_{+0.00512}$ & $0.00282^{+0.00360}_{-0.00468}$ & $-0.00115^{+0.00147}_{-0.00095}$ & $-0.00032^{+0.00007}_{+0.00042}$ & $0.04608^{+0.00015}_{-0.00009}$ \\
 \hline
 ~~$\hat{R}_3|_{\rm Conv.}$~~ & $0.1020^{+0.0471}_{-0.0261}$ & $0.0542^{+0.0089}_{-0.0062}$ & $0.0280^{-0.0176}_{+0.0044}$ & $0.0138^{-0.0214}_{+0.0085}$ & $0.1980^{+0.0170}_{-0.0194}$ \\
 \hline
 ~~$\tilde{R}_3|_{\rm Conv.}$~~ & $0.2030^{-0.0175}_{+0.0226}$ & $0.0374^{+0.0100}_{-0.0151}$ & $0.0019^{+0.0070}_{-0.0077}$ & $-0.0014^{+0.0020}_{-0.0005}$ & $0.2409^{+0.0015}_{-0.0007}$ \\
 \hline
\end{tabular}
\caption{The values of each loop-terms (LO, NLO, N$^2$LO and N$^3$LO) for the four-loop QCD predictions $R_{3}(Q=31.6~{\rm GeV})$, $\hat{R}_{3}(M_{\tau})$ and $\tilde{R}_{3}(M_H)$ using the conventional scale-setting approach. The errors are caused by taking $\mu_r\in [1/2 Q, 2 Q]$ for $R_{3}(Q=31.6~{\rm GeV})$, $\mu_r\in [1{\rm GeV}, 2M_{\tau}]$ for $\hat{R}_{3}(M_{\tau})$, and $\mu_r\in [1/2 M_H, 2 M_H]$ for $\tilde{R}_{3}(M_H)$.}
\label{convorder}
\end{table}
\end{center}
\end{widetext}

In the pQCD calculation, it is helpful to give a reliable prediction of the uncalculated higher-order terms. The Pad\'{e} approximation approach (PAA)~\cite{Basdevant:1972fe, Samuel:1992qg, Samuel:1995jc} provides an effective method to estimate the $(n+1)_{\rm th}$-order coefficient from a given $n_{\rm th}$-order series~\footnote{Another method, which uses the scale-invariant conformal series together with the Bayesian model~\cite{Cacciari:2011ze, Bonvini:2020xeo, Duhr:2021mfd} to provide probabilistic estimates of the unknown higher-orders terms is in preparation.}. Practically, it has been found that the PAA shall become more effective when more loop terms are known. For a pQCD approximant, $\rho(Q)=c_1 a_s+c_2 a_s^2+c_3 a_s^3+c_4 a_s^4+\cdots$, the predicted N$^3$LO-term and the N$^4$LO-term are
\begin{eqnarray}
\rho^{\rm N^3LO}_{[1/1]}&=&\frac{c_3^2} {c_2} a_s^4, \\
\rho^{\rm N^3LO}_{[0/2]}&=&\frac{2c_1 c_2 c_3-c_2^3} {c_1^2} a_s^4, \\
\rho^{\rm N^4LO}_{[1/2]}&=&\frac{2c_2 c_3 c_4-c_3^3-c_1 c_4^2} {c_2^2-c_1 c_3} a_s^5, \\
\rho^{\rm N^4LO}_{[2/1]}&=&\frac{c_4^2} {c_3} a_s^5, \\
\rho^{\rm N^4LO}_{[0/3]}&=&\frac{c_2^4-3c_1 c_2^2 c_3+2c_1^2 c_2 c_4+c_1^2 c_3^2} {c_1^3} a_s^5.
\end{eqnarray}

\begin{table}[htb]
\begin{tabular}{ccc}
\hline
 & $\rm N^3LO$ & $\rm N^4LO$ \\
\hline
 \raisebox {0ex}[0pt]{$R_3|_{\rm Conv.}$}
 & [1/1]:$0.00047^{+0.00000}_{-0.00284}$ & [1/2]:$-0.00002^{+0.00011}_{-0.00010}$ \\
 & - & $[2/1]$:$-0.00009^{+0.00028}_{-0.00000}$ \\
\hline
 \raisebox {0ex}[0pt]{$\hat{R}_3|_{\rm Conv.}$}
 & [1/1]:$0.0145^{+0.0074}_{-0.0128}$ & [1/2]:$+0.0061^{+0.0093}_{-0.0142}$ \\
 & - & $[2/1]$:$+0.0068^{+0.0085}_{-0.0011}$ \\
\hline
 \raisebox {0ex}[0pt]{$\tilde{R}_3|_{\rm Conv.}$}
 & [1/1]:$0.0001^{+0.0016}_{-0.0000}$ & [1/2]:$-0.0006^{+0.0005}_{-0.0000}$ \\
 & - & $[2/1]$:$+0.0010^{+0.0000}_{-0.0016}$ \\
\hline
\end{tabular}
\caption{The preferable diagonal-type PAA predictions of the $\rm N^3LO$ and $\rm N^4LO$ terms of $R_{3}(Q=31.6~{\rm GeV})$, $\hat{R}_{3}(M_\tau)$, and $\tilde{R}_{3}(M_H)$ using the conventional scale-setting approach. The uncertainties correspond to taking $\mu_r\in [1/2 Q, 2 Q]$, $\mu_r\in [1{\rm GeV}, 2M_{\tau}]$ , and $\mu_r\in [1/2 M_H, 2 M_H]$, respectively. } \label{convpaa1}
\end{table}

In Table~\ref{convpaa1}, we give the preferable diagonal-type PAA predictions~\cite{Gardi:1996iq} of the $\rm N^3LO$ and $\rm N^4LO$ terms of $R_{3}(Q=31.6~{\rm GeV})$, $\hat{R}_{3}(M_{\tau})$, and $\tilde{R}_{3}(M_H)$ using the conventional scale-setting approach. Because of the large scale dependence of each loop-terms, the PAA predictions show large scale dependence. The two allowable diagonal-type PAA predictions for $\rm N^4LO$ terms are consistent with each order within errors. Comparing Table \ref{convpaa1} with Table \ref{convorder}, it is noted that the values of the predicted $\rm N^3LO$-terms agree with their exact values within errors. Thus, by employing a perturbative series with enough higher-order terms, the PAA prediction can be reliable.

\subsection{properties using the PMCm approach}

Following the standard PMCm procedures, the nonconformal $\{\beta_i\}$-terms are eliminated by using the RGE recursively, determining the effective $\alpha_s$ at each perturbative order, and resulting in the renormalon-free and scheme independent conformal series (\ref{PMCmseries}).

\begin{table}[htb]
\begin{tabular}{ccccc}
\hline
& ~~~${\hat r}_{1,0}$~~~ & ~~~${\hat r}_{2,0}$~~~ & ~~~${\hat r}_{3,0}$~~~ & ~~~${\hat r}_{4,0}$~~~ \\
 \hline
 $R(Q)$ & $4$ & $29.44$ & $-64.25$ & $-2812.74$ \\
 \hline
 $\hat{R}(M_{\tau})$ & $4$ & $34.33$ & $219.89$ & $1741.15$ \\
 \hline
 $\tilde{R}(M_H)$ & $22.6667$ & $216.356$ & $-8708.09$ & $-110597$ \\
 \hline
\end{tabular}
\caption{Conformal coefficients ${\hat r}_{i,0}$ for $R_{n}(Q=31.6~{\rm GeV})$, $\hat{R}_{n}(M_\tau)$, and $\tilde{R}_{n}(M_H)$, respectively.}
\label{rijn}
\end{table}

We present the conformal coefficients ${\hat r}_{i,0}$ in Table~\ref{rijn}. The PMC scales are of perturbative nature, which leads to the \textit{first kind of residual scale dependence} for PMCm predictions. If the pQCD approximants are known up to four-loop QCD corrections, three PMC scales ($Q_1$, $Q_2$ and $Q_3$) can be determined up to $\rm N^2LL$, $\rm NLL$ and $\rm LL$ order, which are $\{$41.19, 36.85, 168.68$\}$ GeV for $R_{n}(Q=31.6~{\rm GeV})$, $\{$1.26, 0.98, 0.36$\}$ GeV for $\hat{R}_{n}(M_\tau)$, and $\{$62.03, 40.76, 52.76$\}$ GeV for $\tilde{R}_{n}(M_H)$, accordingly. There is no $\{\beta_i\}$-terms to set the scale $Q_4$, the PMCm prediction has the \textit{second kind of residual scale dependence}. As mentioned in Sec.\ref{pmcmsub}, there is the \textit{second kind of residual scale dependence} for PMCm series, and for convenience, we set $Q_4=Q_3$ as the default choice of $Q_4$. A discussion of magnitude of the \textit{second kind of residual scale dependence} by taking some other typical choices of $Q_4$ shall be presented in the end of this subsection. For the scales $>> \Lambda_{\rm QCD}$, we adopt the usual approximate four-loop analytic solution of the RGE to derive the $\alpha_s$ value. Due to the sizable differences between the approximate analytic solution and the exact numerical solution of the RGE at scales below a few GeV~\cite{Wu:2019mky, Workman:2022zbs}, we adopt the exact numerical solution of the RGE to evaluate $R_{\tau}(M_\tau)$ at 1.26 GeV and 0.98 GeV. For the scales close to $\Lambda_{\rm QCD}$, various low-energy models have been suggested in the literature; a detailed comparison of various low-energy models can be found in Ref.\cite{Zhang:2014qqa}. For definiteness, we will adopt the Massive Perturbation Theory (MPT) model~\cite{Shirkov:2012ux} to evaluate $R_\tau(M_\tau)$ at $Q_3=0.36$ GeV, which gives $\alpha_s|_{\rm MPT}^{ \xi=10\pm2}(0.36)=0.559^{+0.042}_{-0.032}$, where $\xi$ is the parameter in the MPT model.

\begin{table}[htb]
\begin{tabular}{ccccccc}
\hline
& $n=1$ & $n=2$ & $n=3$ & $\kappa_1$ & $\kappa_2$ & $\kappa_3$ \\
 \hline
 $R_n|_{\rm PMCm}$ & $0.04735$ & $0.04640$ & $0.04610$ & $6.9\%$ & $2.0\%$ & $0.6\%$ \\
 \hline
 $\hat{R}_n|_{\rm PMCm}$ & $0.2133$ & $0.1991$ & $0.2087$ & $139.4\%$ & $6.7\%$ & $4.8\%$ \\
 \hline
 $\tilde{R}_n|_{\rm PMCm}$ & $0.2481$ & $0.2402$ & $0.2400$ & $22.0\%$ & $3.2\%$ & $0.1\%$ \\
 \hline
\end{tabular}
\caption{Results for $R_{n}(Q=31.6~{\rm GeV})$, $\hat{R}_{n}(M_{\tau})$, $\tilde{R}_{n}(M_H)$ up to four-loop QCD corrections using the PMCm scale-setting approach. The renormalization scale is set as $Q$, $M_\tau$, or $M_H$, respectively. }
\label{pmcm}
\end{table}

We present the results of $R_{n}(Q=31.6~{\rm GeV})$, $\hat{R}_{n}(M_\tau)$, and $\tilde{R}_{n}(M_H)$ up to four-loop QCD corrections using the PMCm scale-setting approach in Table~\ref{pmcm}. Table~\ref{pmcm} shows that the PMCm predictions generally have close behavior to the central predictions under conventional scale-setting procedures, especially when more loop terms are known. This is because that when the renormalization scale of the conventional series is set as the one to eliminate the large logs, the divergent renormalon terms may also be simultaneously removed, since the $\{\beta_i\}$-terms are always accompanied by the log-terms.

\begin{center}
\begin{table*}[htb]
\begin{tabular}{cccccc}
\hline
 & ~~~$\rm LO$~~~ & ~~~$\rm NLO$~~~ & ~~~$\rm N^2LO$~~~ & ~~~$\rm N^3LO$~~~ & ~~~$\rm Total$~~~ \\
 \hline
 ~~~$R_3|_{\rm PMCm}$~~~ & $0.04267^{+0.00003}_{-0.00001}$ & $0.00349^{+0.00003}_{-0.00004}$ & $-0.00004$ & $-0.00002$ & $0.04610^{+0.00006}_{-0.00005}$ \\
 \hline
 $\hat{R}_3|_{\rm PMCm}$ & $0.1272^{+0.0062}_{-0.0090}$ & $0.0553^{+0.0147}_{-0.0178}$ & $0.0194$ & $0.0068$ & $0.2087^{+0.0209}_{-0.0268}$ \\
 \hline
 $\tilde{R}_3|_{\rm PMCm}$ & $0.2258$ & $0.0247^{+0.0001}_{-0.0001}$ & $-0.0093$ & $-0.0012$ & $0.2400^{+0.0001}_{-0.0001}$ \\
 \hline
\end{tabular}
\caption{The values of each loop-terms (LO, NLO, N$^2$LO and N$^3$LO) for the four-loop QCD predictions $R_{3}(Q=31.6~{\rm GeV})$, $\hat{R}_{3}(M_{\tau})$ and $\tilde{R}_{3}(M_H)$ using the PMCm scale-setting approach. The uncertainties correspond to taking $\mu_r\in [1/2 Q, 2 Q]$ for $R_{3}(Q=31.6~{\rm GeV})$, $\mu_r\in [1{\rm GeV}, 2M_{\tau}]$ for $\hat{R}_{3}(M_{\tau})$, and $\mu_r\in [1/2 M_H, 2 M_H]$ for $\tilde{R}_{3}(M_H)$.}
\label{pmcmorder}
\end{table*}
\end{center}

We present the values of each loop terms for the four-loop predictions $R_{3}(Q=31.6~{\rm GeV})$, $\hat{R}_{3}(M_{\tau})$, and $\tilde{R}_{3}(M_H)$ using the PMCm scale-setting approach in Table~\ref{pmcmorder}~\footnote{ By using the RGE recursively, one can get correct $\alpha_s$-value and achieve a well-matching of $\alpha_s$ to its coefficients at the same perturbative order. However such treatment is a kind of $\{\beta_i\}$-resummation, and the resultant PMC series is no longer a usual fixed-order series. Thus the function of Table~\ref{pmcmorder} is to show its own perturbative behavior.}. The relative importance among the LO-terms, the NLO-terms, the N$^2$LO-terms and the N$^3$LO-terms for those approximants are
\begin{eqnarray}
&& 1:+0.0818: -0.0009: -0.0005, \;(\mu_r=Q) \\
&& 1:+0.4347: +0.1525: +0.0535, \;(\mu_r=M_\tau) \\
&& 1:+0.1094: -0.0412: -0.0053. \;(\mu_r=M_H)
\end{eqnarray}
Table \ref{pmcmorder} shows that there are residual scale dependence of $R_{3}(Q=31.6~{\rm GeV})$ and $\tilde{R}_{3}(M_H)$ for the LO and NLO terms, which are quite small (e.g. the errors are only about $\pm 0.1\%$ of the LO-terms and $\pm 0.04\%$ of the NLO-terms, respectively), which are smaller than the corresponding ones under conventional scale-setting approach. Although the residual scale dependence of $\hat{R}_{3}(M_{\tau})$ is sizable -- approximately $\pm 10\%$ and is comparable to the conventional scale dependence. Here the large residual scale dependence of $\hat{R}_{3}(M_{\tau})$ is reasonable, caused by the poor pQCD convergence for the PMC scales at higher orders and the uncertainties of the $\alpha_s$-running behavior in the low-energy region.

\begin{center}
\begin{table*}[htb]
\begin{tabular}{cccccc}
\hline
 & $Q_4=2Q_3$ & $Q_4=\frac{1}{2}Q_3$ & $Q_4=Q_1$ & $Q_4=Q_2$ & $Q_4={(Q_1+Q_2+Q_3)}/{3}$ \\
 \hline
 $R_3|_{\rm PMCm}$ & $0.04611$ & $0.04610$ & $0.04608$ & $0.04608$ & $0.04610$ \\
 \hline
 $\hat{R}_3|_{\rm PMCm}$ & $0.2054^{+0.0054}_{-0.0036}$ & $0.2105^{+0.0088}_{-0.0053}$ & $0.2033^{+0.0049}_{-0.0032}$ & $0.2041^{+0.0050}_{-0.0033}$ & $0.2046^{+0.0051}_{-0.0034}$ \\
 \hline
 $\tilde{R}_3|_{\rm PMCm}$ & $0.2404$ & $0.2392$ & $0.2401$ & $0.2398$ & $0.2400$ \\
 \hline
\end{tabular}
\caption{The four-loop pQCD predictions of $R_{3}(Q=31.6~{\rm GeV})$, $\hat{R}_{3}(M_{\tau})$ and $\tilde{R}_{3}(M_H)$ under some other choices of the undetermined $Q_4$ as an estimation of the \textit{second kind of residual scale dependence} under the PMCm scale-setting approach. The uncertainties for $\hat{R}_{3}(M_{\tau})$ is obtained by changing the MPT model parameter $\xi=10\pm2$.}
\label{Q4effect}
\end{table*}
\end{center}

As a final remark, we discuss the possible magnitudes of the \textit{second kind of residual scale dependence} under the PMCm scale-setting approach by taking some other typical choices of $Q_4$, e.g. $2 Q_3$, $(1/2) Q_3$, $Q_1$, $Q_2$ and $(Q_1+Q_2+Q_3)/3$, which also ensures the scheme independence of the PMCm series. Table \ref{Q4effect} shows that the \textit{second kind of residual scale dependence} of $R_3$, $\hat{R}_3$ and $\tilde{R}_3$ are $(^{+0.00001}_{-0.00002})$, $(^{+0.0018}_{-0.0054})$ and $(^{+0.0004}_{-0.0008})$, respectively. It shows that those choices shall change the magnitudes of $R_3$, $\hat{R}_3$ and $\tilde{R}_3$ at the default choice of $Q_4=Q_3$ by about $\pm 0.04\%$, $\pm 2.6\%$ and $\pm 0.3\%$, respectively. For $\hat{R}_{3}(M_{\tau})$, another uncertainty caused by the MPT model parameter $\xi=10\pm2$ is $\hat{R}_3|_{\rm PMCm}=0.2087^{+0.0072}_{-0.0046}$ in the choice of $Q_4=Q_3$. We also give the numerical results under some other choices of the undetermined $Q_4$ in Table \ref{Q4effect}. It shows that the MPT model parameter $\xi=10\pm2$ will lead to $\sim3\%$ uncertainties. These uncertainties caused by the small scale $Q_3$ and undetermined scale $Q_4$ indicate that we still need a more appropriate scale-setting approach to suppress the theoretical uncertainties.

\subsection{properties using the PMCs approach}

The PMCs approach provides a method to suppress the residual scale dependence. Applying the standard PMCs scale-setting procedures, we obtain an overall effective $\alpha_s$, and accordingly, an overall effective scale ($Q_*$) for $R_{n}(Q=31.6~{\rm GeV})$, $\hat{R}_{n}(M_{\tau})$, and $\tilde{R}_{n}(M_H)$, respectively. If they are known up to two-loop, three-loop, and four-loop levels, the PMC scale $Q_*$ can be determined up to ${\rm LL}$, ${\rm NLL}$ and ${\rm N^2LL}$ accuracies, respectively. That is, for $n=1,2,3$, we have
\begin{eqnarray}
Q_{*}|_{e^+e^-} &=& \{35.36, 39.49, 40.12\} \; {\rm GeV}, \\
Q_{*}|_{\tau} &=& \{0.90, 1.06, 1.07 \} \; {\rm GeV}, \\
Q_{*}|_{H\to b\bar{b}} &=& \{60.94, 56.51, 58.80\} \; {\rm GeV}.
\end{eqnarray}
Their magnitudes become more precise as one includes more loop terms, and the difference between the two nearby values becomes smaller and smaller when more loop terms are included, e.g. the ${\rm N^2LL}$ scales only shift about $2\%-4\%$ to the ${\rm NLL}$ ones. Since these PMCs scales are numerically sizable, one avoids confronting the possibly small scale problem at certain perturbative orders of the multi-scale-setting approaches such as PMCm, PMC$_\infty$ and PMCa.

\begin{table}[htb]
\begin{tabular}{ccccccc}
\hline
& $n=1$ & $n=2$ & $n=3$ & $\kappa_1$ & $\kappa_2$ & $\kappa_3$ \\
 \hline
 $R_n|_{\rm PMCs}$ & $0.04735$ & $ 0.04629$ & $ 0.04613$ & $6.9\%$ & $2.2\%$ & $0.3\%$ \\
 \hline
 $\hat{R}_n|_{\rm PMCs}$ & $0.2136$ & $ 0.1997$ & $ 0.2064$ & $139.7\%$ & $6.5\%$ & $3.4\%$ \\
 \hline
 $\tilde{R}_n|_{\rm PMCs}$ & $0.2481$ & $ 0.2424$ & $ 0.2398$ & $22.0\%$ & $2.3\%$ & $1.1\%$ \\
 \hline
\end{tabular}
\caption{Results for $R_{n}(Q=31.6~{\rm GeV})$, $\hat{R}_{n}(M_{\tau})$, $\tilde{R}_{n}(M_H)$ up to four-loop QCD corrections using the PMCs scale-setting approach, which are independent to any choice of renormalization scale.}
\label{pmcs}
\end{table}

\begin{table}[htb]
\begin{tabular}{cccccc}
\hline
 & $\rm LO$ & $\rm NLO$ & $\rm N^2LO$ & $\rm N^3LO$ & $\rm Total$ \\
 \hline
 $R_3|_{\rm PMCs}$ & $0.04287$ & $0.00338$ & $-0.00008$ & $-0.00004$ & $0.04613$ \\
 \hline
 $\hat{R}_3|_{\rm PMCs}$ & $0.1465$ & $0.0460$ & $0.0108$ & $0.0031$ & $0.2064$ \\
 \hline
 $\tilde{R}_3|_{\rm PMCs}$ & $0.2278$ & $0.0219$ & $-0.0088$ & $-0.0011$ & $0.2398$ \\
 \hline
\end{tabular}
\caption{The values of each loop-terms (LO, NLO, N$^2$LO or N$^3$LO) for the four-loop predictions $R_{3}(Q=31.6~{\rm GeV})$, $\hat{R}_{3}(M_{\tau})$, $\tilde{R}_{3}(M_H)$ using the PMCs scale-setting approach.}
\label{pmcsorder}
\end{table}

\begin{figure}[htb]
\includegraphics[width=0.5\textwidth]{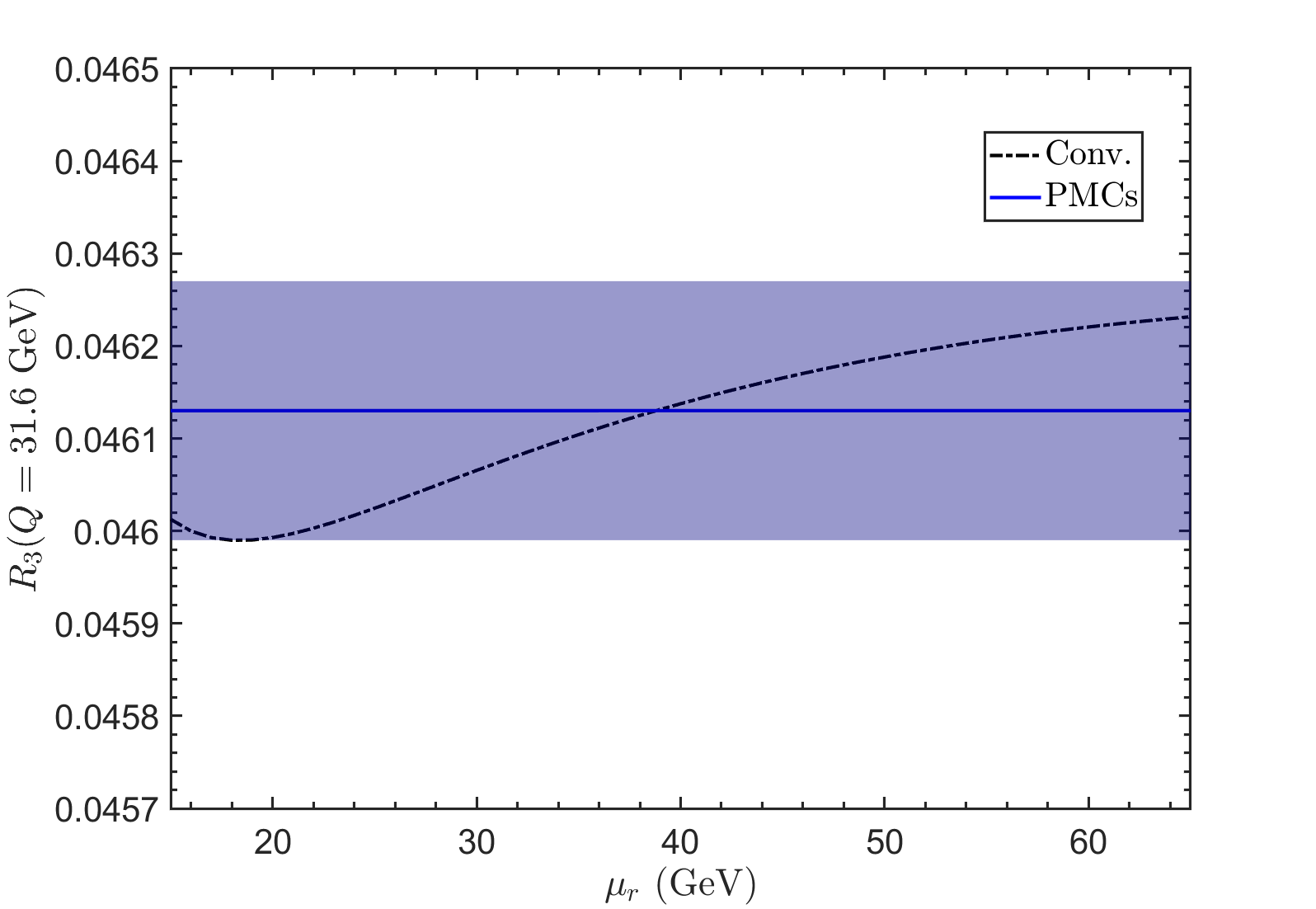}
\caption{The renormalization scale dependence of the four-loop prediction $R_{3}(Q=31.6~{\rm GeV})$ using the conventional and PMCs scale-setting procedures. The band represents a conservative estimation (\ref{PMCsfirstres1}) of the \textit{first kind of residual scale dependence} under the PMCs.}  \label{urdepen11}
\end{figure}

\begin{figure}[htb]
\includegraphics[width=0.5\textwidth]{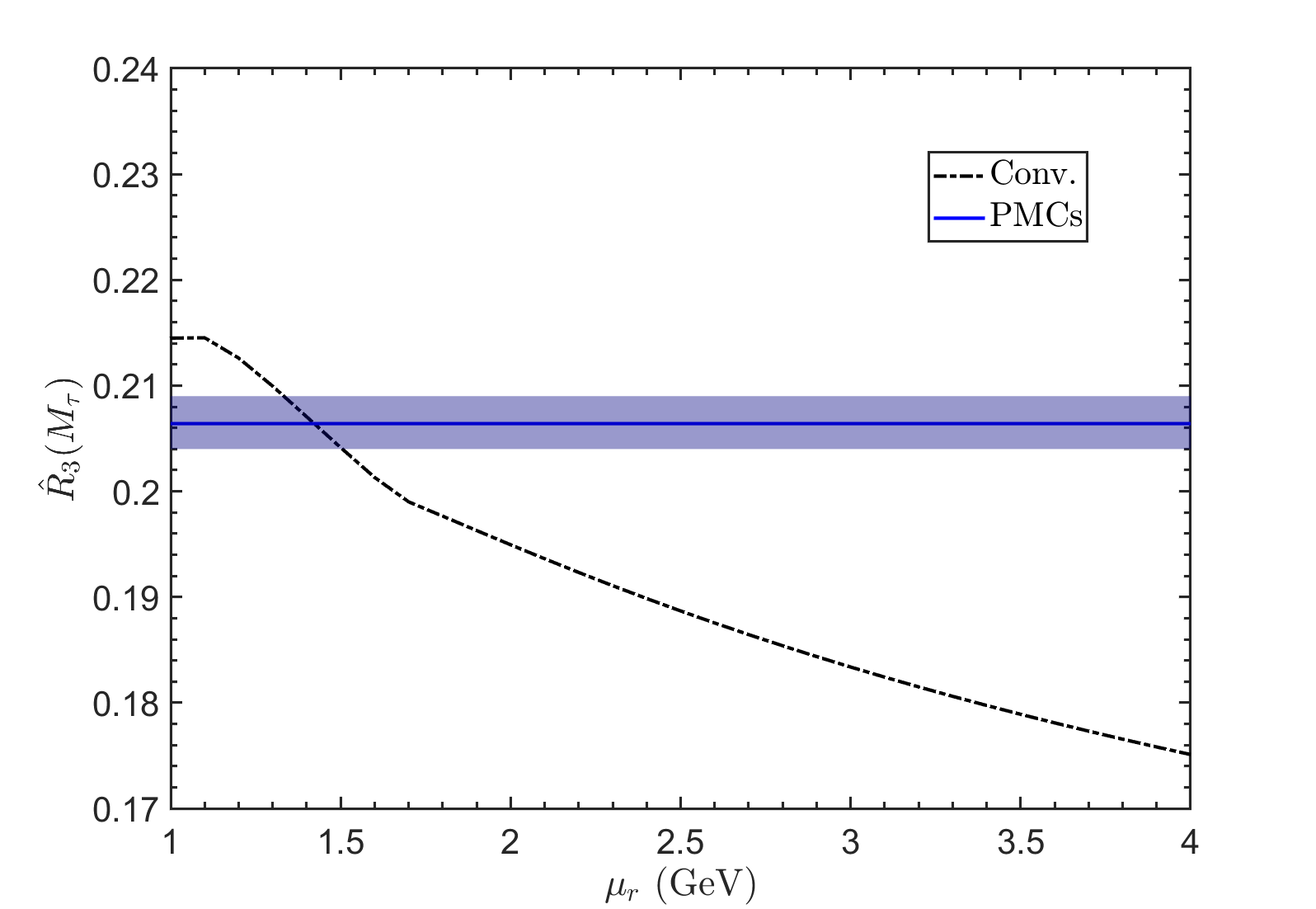}
\caption{The renormalization scale dependence of the four-loop prediction $\hat{R}_{3}(M_{\tau})$ using the conventional and PMCs scale-setting procedures. The band represents a conservative estimation (\ref{PMCsfirstres2}) of the \textit{first kind of residual scale dependence} under the PMCs.} \label{urdepen21}
\end{figure}

\begin{figure}[htb]
\includegraphics[width=0.5\textwidth]{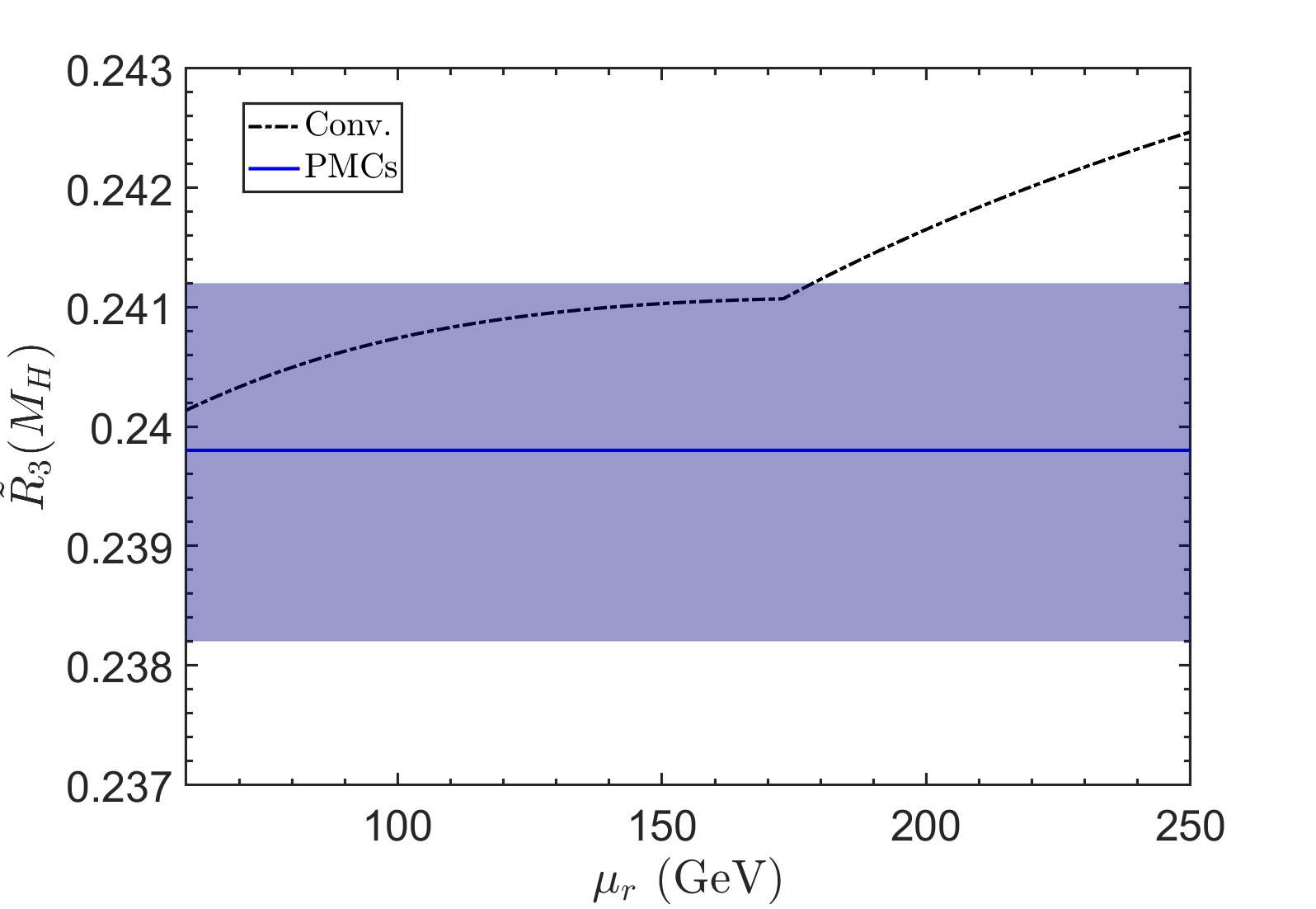}
\caption{The renormalization scale dependence of the four-loop prediction $\tilde{R}_{3}(M_H)$ using the conventional and PMCs scale-setting procedures. The band represents a conservative estimation (\ref{PMCsfirstres3}) of the \textit{first kind of residual scale dependence} under the PMCs.} \label{urdepen31}
\end{figure}

We present the results of $R_{n}(Q=31.6~{\rm GeV})$, $\hat{R}_{n}(M_{\tau})$, and $\tilde{R}_{n}(M_H)$ up to four-loop QCD corrections using the PMCs scale-setting approach in Table~\ref{pmcs}. We also present the values of each loop-term for the four-loop approximants $R_{3}(Q=31.6~{\rm GeV})$, $\hat{R}_{3}(M_{\tau})$, and $\tilde{R}_{3}(M_H)$ using the PMCs scale-setting approach in Table~\ref{pmcsorder}. The relative importance among the LO-terms, the NLO-terms, the N$^2$LO-terms and the N$^3$LO-terms for those approximants are
\begin{eqnarray}
&& 1 : +0.0788 : -0.0019 : -0.0009, \\
&& 1 : +0.3140 : +0.0737: +0.0212, \\
&& 1 : +0.0961 : -0.0386 : -0.0048.
\end{eqnarray}
which are comparable to the convergent behaviors of the PMCm series and more convergent than conventional predictions. Moreover, sizable residual scale dependence of $\hat{R}_{3}(M_{\tau})$ appeared in Table~\ref{pmcmorder} has been eliminated by using the PMCs procedure. Thus the PMCs approach, which requires a much simpler analysis, can be adopted as a reliable substitute for the basic PMCm approach for setting the renormalization scales for high-energy processes with small residual scale dependence. As a conservative estimation of the \textit{first kind of residual scale dependence}, we take the magnitude of its last known term as the unknown N$^3$LL-term, e.g. $\pm( |S_2 a^2_s(Q_*)|)$ for $R_3(Q=31.6~{\rm GeV})$, $ {\hat R}_3(M_{\tau})$ and ${\tilde R}_3(M_H)$. Then we obtain
\begin{eqnarray}
  R_3(Q=31.6~{\rm GeV}) &=&0.04613\pm 0.00014, \label{PMCsfirstres1} \\
 {\hat R}_3(M_{\tau}) &=&0.2064^{+0.0026}_{-0.0024}, \label{PMCsfirstres2} \\
 {\tilde R}_3(M_H) &=&0.2398^{+0.0014}_{-0.0016}, \label{PMCsfirstres3}
\end{eqnarray}
which show the \textit{first kind of residual scale dependence} are about $\pm0.3\%$, $\pm1.3\%$ and $\pm0.7\%$~\footnote{Since the N$^2$LL-accuracy $\ln Q^2_*/Q^2$-series of those pQCD approximants already show good perturbative behavior, it is found that by using the PAA predicted N$^3$LL-term, e.g. $\pm |{S^2_2}/{S_1} a^3_s(Q_*)|$, to do the estimation, one can obtain smaller \textit{first kind of residual scale dependence} than the ones listed in Eqs.(\ref{PMCsfirstres1}, \ref{PMCsfirstres2}, \ref{PMCsfirstres3}), which are $\pm 0.00002$, $\left(^{+0.0002}_{-0.0003}\right)$, and $\left(^{+0.0007}_{-0.0009}\right)$, respectively.}. More explicitly, we show the conservative estimation of the \textit{first kind of residual scale dependence} under the PMCs in Figs.~\ref{urdepen11}, \ref{urdepen21}, and \ref{urdepen31}, respectively.

The conformal PMCs series is scheme and scale independent; it thus provides a reliable basis for estimating the effect of unknown higher-order contributions. At present, there is no way to use a series with different effective $\alpha_s(Q_i)$ at different orders; and if there were any, its effectiveness would also be greatly affected by the possibly large residual scale dependence. Thus we shall not use PMCm, PMC$_\infty$ and PMCa series to estimate the contribution of the unknown terms. As for the PMCs series, with an overall effective $\alpha_s(Q_*)$, we can directly use the PAA~\cite{Du:2018dma}.

\begin{table}[htb]
\begin{tabular}{ccc}
\hline
& ~~~${\hat r}_{4,0}$~~~ & ~~~${\hat r}_{5,0}$~~~  \\
 \hline
 $R(Q)$ & [0/2]:$-2541.35$ & [0/3]:$-181893$ \\
 \hline
 $\hat{R}(M_{\tau})$ & [0/2]:$1245.3$ & [0/3]:$15088.6$ \\
 \hline
 $\tilde{R}(M_H)$ & [0/2]:$-185951$ & [0/3]:$3802450$ \\
 \hline
\end{tabular}
\caption{The preferable [0/$n$-1]-type PAA predictions of conformal coefficients ${\hat r}_{4,0}$ and ${\hat r}_{5,0}$ for $R_{n}(Q=31.6~{\rm GeV})$, $\hat{R}_{n}(M_\tau)$, and $\tilde{R}_{n}(M_H)$, respectively.}
\label{rijn2}
\end{table}

\begin{table}[htb]
\begin{tabular}{ccc}
\hline
 & ~~~$\rm N^3LO$~~~ & ~~~$\rm N^4LO$~~~ \\
\hline
 \raisebox {0ex}[0pt]{~~~$R_3|_{\rm PMCs}$~~~}
 & [0/2]:$-0.00003$ & [0/3]:$-0.000003$ \\
\hline
 \raisebox {0ex}[0pt]{$\hat{R}_3|_{\rm PMCs}$}
 & [0/2]:$+0.0022$ & [0/3]:$+0.0010$ \\
\hline
 \raisebox {0ex}[0pt]{$\tilde{R}_3|_{\rm PMCs}$}
 & [0/2]:$-0.0019$ & [0/3]:$+0.0004$ \\
\hline
\end{tabular}
\caption{The preferable [0/$n$-1]-type PAA predictions of the $\rm N^3LO$ and $\rm N^4LO$ terms of $R_{3}(Q=31.6~{\rm GeV})$, $\hat{R}_{3}(M_{\tau})$, and $\tilde{R}_{3}(M_H)$ using the PMCs scale-setting approach.} \label{pmcpaa1}
\end{table}

We present the preferable [0/$n$-1]-type PAA predictions for the PMCs series of $R_{3}(Q=31.6~{\rm GeV})$, $\hat{R}_{3}(M_{\tau})$, and $\tilde{R}_{3}(M_H)$ in Table~\ref{rijn2} and \ref{pmcpaa1}. Table~\ref{rijn2} shows the predicted $\rm N^3LO$ and $\rm N^4LO$ conformal coefficients ${\hat r}_{4,0}$ and ${\hat r}_{5,0}$. It is noted that the predicted ${\hat r}_{4,0}$ values are close to the exact values shown in Table~\ref{rijn}, and those known conformal coefficients do not change when more loop terms are known. To obtain the final numerical result, we need to combine the coefficients ${\hat r}_{4,0}$ and ${\hat r}_{5,0}$ with the effective $\alpha_s(Q_*)$ with corresponding orders. Table~\ref{pmcpaa1} shows their numerical results, those values shall be very slightly changed for a more accurate $Q_*$, since the ${\rm N^2LL}$ accuracy $Q_*$ is already changed from ${\rm NLL}$ one by less than $\sim 5\%$.

\subsection{properties using the PMC$_\infty$ and the PMCa approaches}

Since the PMC$_\infty$ and PMCa methods are equivalent, we shall only give the results of PMC$_\infty$.

Following the standard PMC$_\infty$ procedures, we calculate $R_{n}(Q=31.6~{\rm GeV})$, $\hat{R}_{n}(M_{\tau})$, and $\tilde{R}_{n}(M_H)$ up to four-loop QCD corrections. The perturbative coefficients ($\mathcal{C}_{i=(1,\cdots,4),\rm{IC}}$) are exactly the same as those of PMCm and PMCs conformal coefficients (${\hat r}_{i=(1,\cdots,4),0}$). As shown by Eqs.(\ref{Infq1}, \ref{Infq2}, \ref{Infq3}), the PMC$_\infty$ scales are definite and have no perturbative nature, which are free of renormalization scale ambiguities, and do not have the \textit{first kind of residual scale dependence}. Using the four-loop QCD corrections, we can determine their first three scales, e.g.
\begin{eqnarray}
\{\mu_{\rm I},\mu_{\rm II}, \mu_{\rm III}\}|_{e^+e^-} &=& \{35.36, 71.11, 0.003\} \;{\rm GeV}, \\
\{\mu_{\rm I},\mu_{\rm II}, \mu_{\rm III}\}|_{\tau} &=& \{0.90, 1.16, 1.82\} \;{\rm GeV}, \\
\{\mu_{\rm I},\mu_{\rm II}, \mu_{\rm III}\}|_{H\to b\bar{b}} &=& \{60.94, 41.24, 46.44\} \;{\rm GeV}.
\end{eqnarray}
For the case of $R_{3}(Q=31.6~{\rm GeV})$, its third scale $\mu_{\rm III}=0.003$ GeV is quite small, and we adopt the above mentioned MPT model to estimate its contribution, which gives $\alpha_s|_{\rm MPT}(0.003)=0.606$. As mentioned in Sec.\ref{subsec-c}, since the ${\rm N^4LO}$ coefficient ${\cal C}_5$ is unknown, the fourth scale $\mu_{\rm IV}$ can not be fixed and we keep its central value as the original one ($\mu_r$) and use $\mu_{\rm IV}\in [\mu_r/2, 2\mu_r]$ to ascertain its \textit{second kind of residual scale dependence}.

\begin{table}[htb]
\begin{tabular}{ccccccc}
\hline
& $n=1$ & $n=2$ & $n=3$ & $\kappa_1$ & $\kappa_2$ & $\kappa_3$ \\
 \hline
 $R_n|_{\rm PMC_\infty}$ & $0.04750$ & $0.04652$ & $0.03937$ & $7.3\%$ & $2.1\%$ & $15.4\%$ \\
 \hline
 $\hat{R}_n|_{\rm PMC_\infty}$ & $0.1805$ & $0.2112$ & $0.2199$ & $102.6\%$ & $17.0\%$ & $4.1\%$ \\
 \hline
 $\tilde{R}_n|_{\rm PMC_\infty}$ & $0.2438$ & $0.2448$ & $0.2405$ & $19.9\%$ & $0.4\%$ & $1.8\%$ \\
 \hline
\end{tabular}
\caption{Results for $R_{n}(Q=31.6~{\rm GeV})$, $\hat{R}_{n}(M_{\tau})$, $\tilde{R}_{n}(M_H)$ up to four-loop QCD corrections using the PMC$_\infty$ scale-setting approach. For each case, the undetermined PMC$_\infty$ scale of the highest order terms is set as $Q$, $M_\tau$, and $M_H$, respectively.}
\label{pmcinf}
\end{table}

\begin{center}
\begin{table*}[htb]
\begin{tabular}{cccccc}
\hline
 ~~~ ~~~ & ~~~$\rm LO$~~~ & ~~~$\rm NLO$~~~ & ~~~$\rm N^2LO$~~~ & ~~~$\rm N^3LO$~~~ & ~~~$\rm Total$~~~ \\
 \hline
 $R_3|_{\rm PMC_\infty}$ & $0.04383$ & $0.00280$ & $-0.00722$ & $-0.00004^{+0.00001}_{-0.00004}$ & $0.03937^{+0.00001}_{-0.00004}$ \\
 \hline
 $\hat{R}_3|_{\rm PMC_\infty}$ & $0.1761$ & $0.0396$ & $0.0035$ & $0.0007^{+0.0034}_{-0.0005}$ & $0.2199^{+0.0034}_{-0.0005}$ \\
 \hline
 $\tilde{R}_3|_{\rm PMC_\infty}$ & $0.2265$ & $0.0246$ & $-0.0099$ & $-0.0007^{+0.0002}_{-0.0004}$ & $0.2405^{+0.0002}_{-0.0004}$ \\
 \hline
\end{tabular}
\caption{The values of each loop-terms (LO, NLO, N$^2$LO or N$^3$LO) for the four-loop predictions $R_{3}(Q=31.6~{\rm GeV})$, $\hat{R}_{3}(M_{\tau})$, and $\tilde{R}_{3}(M_H)$ using the PMC$_\infty$ scale-setting approach. The errors, representing the \textit{second kind of residual scale dependence}, are estimated by varying the undetermined PMC$_\infty$ scale $\mu_{\rm IV}$ within the region of $[1/2Q, 2Q]$ for $R_{3}(Q=31.6~{\rm GeV})$, $[1{\rm GeV}, 2M_{\tau}]$ for $\hat{R}_{3}(M_{\tau})$, and $[1/2M_H, 2M_H]$ for $\tilde{R}_{3}(M_H)$.}
\label{pmcinforder}
\end{table*}
\end{center}

We present the results of $R_{n}(Q=31.6~{\rm GeV})$, $\hat{R}_{n}(M_{\tau})$, and $\tilde{R}_{n}(M_H)$ up to four-loop QCD corrections using the PMC$_\infty$ scale-setting approach in Table~\ref{pmcinf}. For the cases of $R_{n}(Q=31.6~{\rm GeV})$ and $\tilde{R}_{n}(M_H)$, we have $\kappa_2 < \kappa_3$, indicating the \textit{second kind of residual scale dependence} is sizable for those two quantities which largely affects the magnitude of the lower-order series. When one has enough higher-order terms, the residual scale dependence is highly suppressed due to the more convergent renormalon-free series. For example, we present the values of each loop-terms (LO, NLO, N$^2$LO or N$^3$LO) for the four-loop predictions $R_{3}(Q=31.6~{\rm GeV})$, $\hat{R}_{3}(M_{\tau})$, and $\tilde{R}_{3}(M_H)$ in Table~\ref{pmcinforder}. At the four-loop level, the PMC$_\infty$ series already exhibits convergent behavior. As shown in Table~\ref{pmcinforder}, the relative importance among the LO-terms, the NLO-terms, the N$^2$LO-terms and the N$^3$LO-terms for those approximants are
\begin{eqnarray}
&& 1:+0.0639: -0.1647: -0.0009, \;(\mu_r=Q) \\
&& 1:+0.2249: +0.0199: +0.0040, \;(\mu_r=M_\tau) \\
&& 1:+0.1086: -0.0437: -0.0031. \;(\mu_r=M_H)
\end{eqnarray}
This perturbative behavior is similar to the predictions of PMCm and PMCs, except for $R_{3}(Q=31.6~{\rm GeV})$, which due to a much smaller scale $\mu_{\rm III}$ leads to quite large N$^2$LO-terms.

\subsection{A comparison of the renormalization scale dependence of the various PMC approaches }

\begin{figure}[htb]
\includegraphics[width=0.5\textwidth]{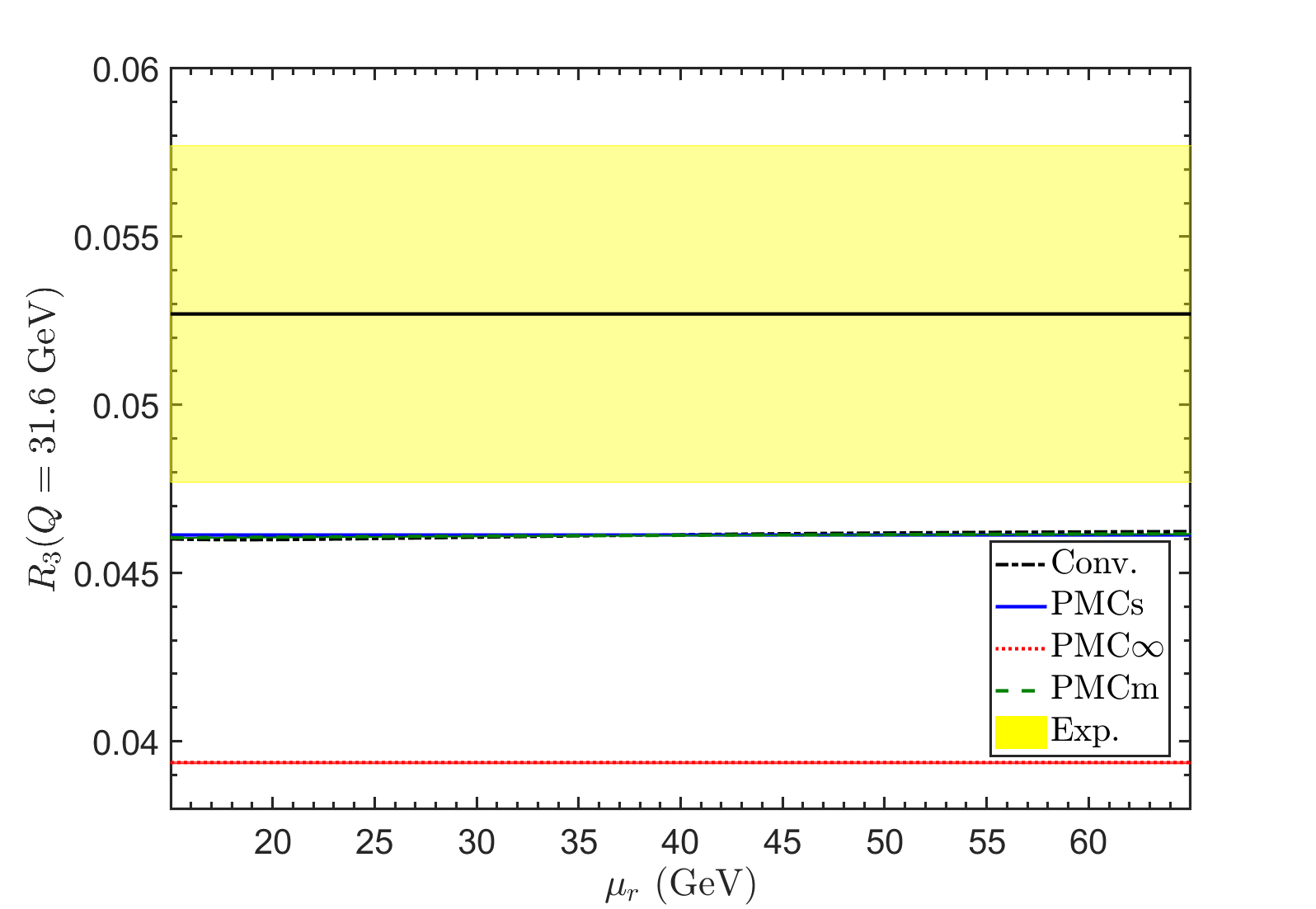}
\caption{The renormalization scale dependence of the four-loop prediction $R_{3}(Q=31.6~{\rm GeV})$ using the conventional, PMCm, PMCs and PMC$_\infty$ scale-setting procedures. The green/lighter band represents the \textit{second kind of residual scale dependence} under the PMCm, which is obtained by changing the undetermined $Q_4$ to be $2 Q_3$, $(1/2)Q_3$, $Q_1$, $Q_2$ and $(Q_1+Q_2+Q_3)/3$, respectively. The red/darker band represents the \textit{second kind of residual scale dependence} under the PMC$_\infty$, which is obtained by varying the undetermined PMC$_\infty$ scale $\mu_{\rm IV}$ within the region of $[1/2Q, 2Q]$. The experimental result $R^{\rm Exp.}(Q=31.6~{\rm GeV})=0.0527\pm 0.0050$ is extracted from $\frac{3}{11}R_{e^+ e^-}^{\rm Exp.}=1.0527 \pm0.0050$~\cite{Marshall:1988ri}. } \label{urdepen1}
\end{figure}

\begin{figure}[htb]
\includegraphics[width=0.5\textwidth]{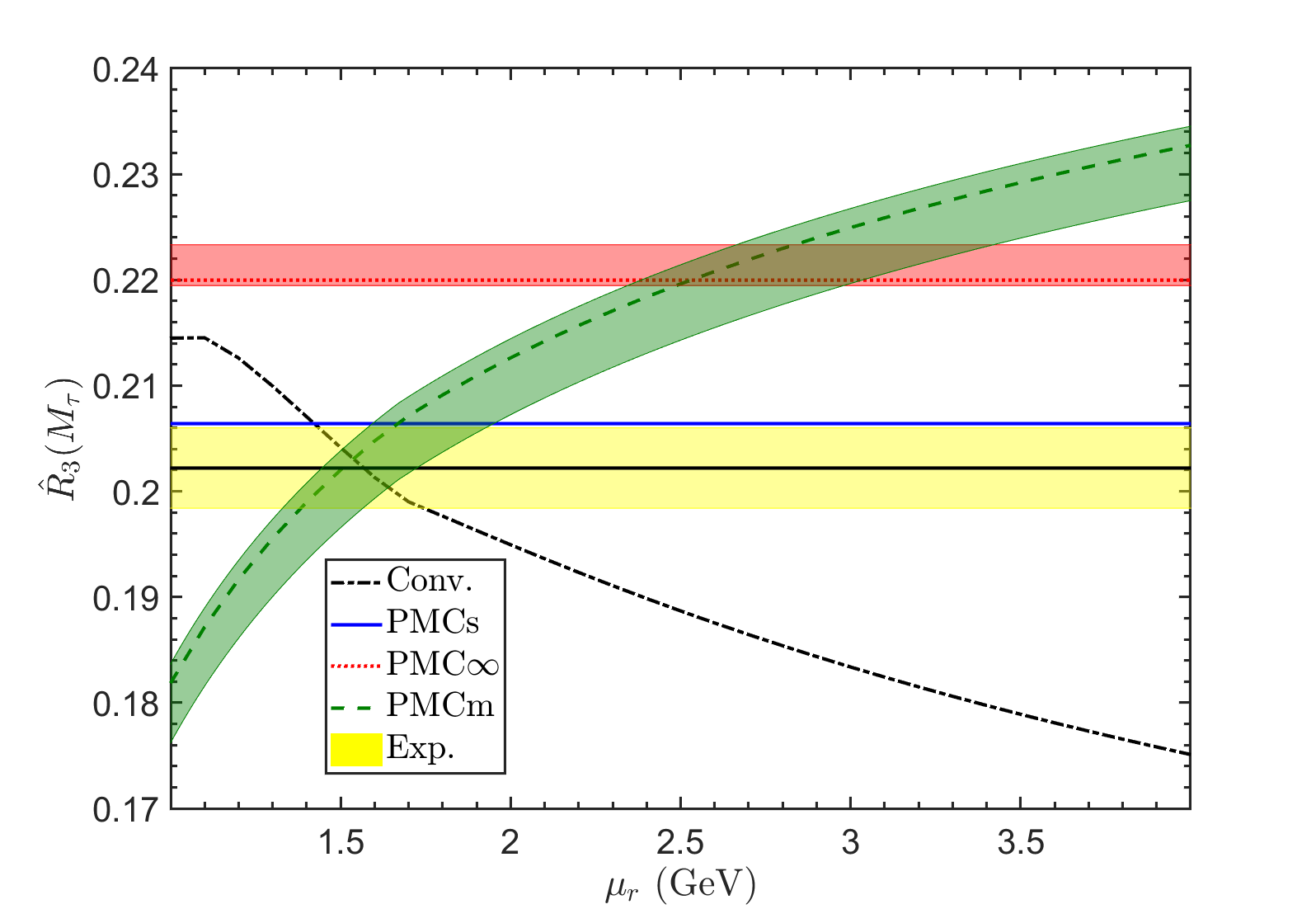}
\caption{The renormalization scale dependence of the four-loop prediction $\hat{R}_{3}(M_{\tau})$ using the conventional, PMCm, PMCs and PMC$_\infty$ scale-setting procedures. The green/lighter band represents the \textit{second kind of residual scale dependence} under the PMCm, which is obtained by changing the undetermined $Q_4$ to be $2 Q_3$, $(1/2)Q_3$, $Q_1$, $Q_2$ and $(Q_1+Q_2+Q_3)/3$, respectively. The red/darker band represents the \textit{second kind of residual scale dependence} under the PMC$_\infty$, which is obtained by varying the undetermined PMC$_\infty$ scale $\mu_{\rm IV}$ within the region of $[1{\rm GeV}, 2M_{\tau}]$. The experimental result $\hat{R}^{\rm Exp.}(M_{\tau})=0.2022^{+0.0038}_{-0.0038}$ is extracted from $R_{\tau}^{\rm Exp.}(M_{\tau})=3.475 \pm0.011$~\cite{Davier:2013sfa}. } \label{urdepen2}
\end{figure}

\begin{figure}[htb]
\includegraphics[width=0.5\textwidth]{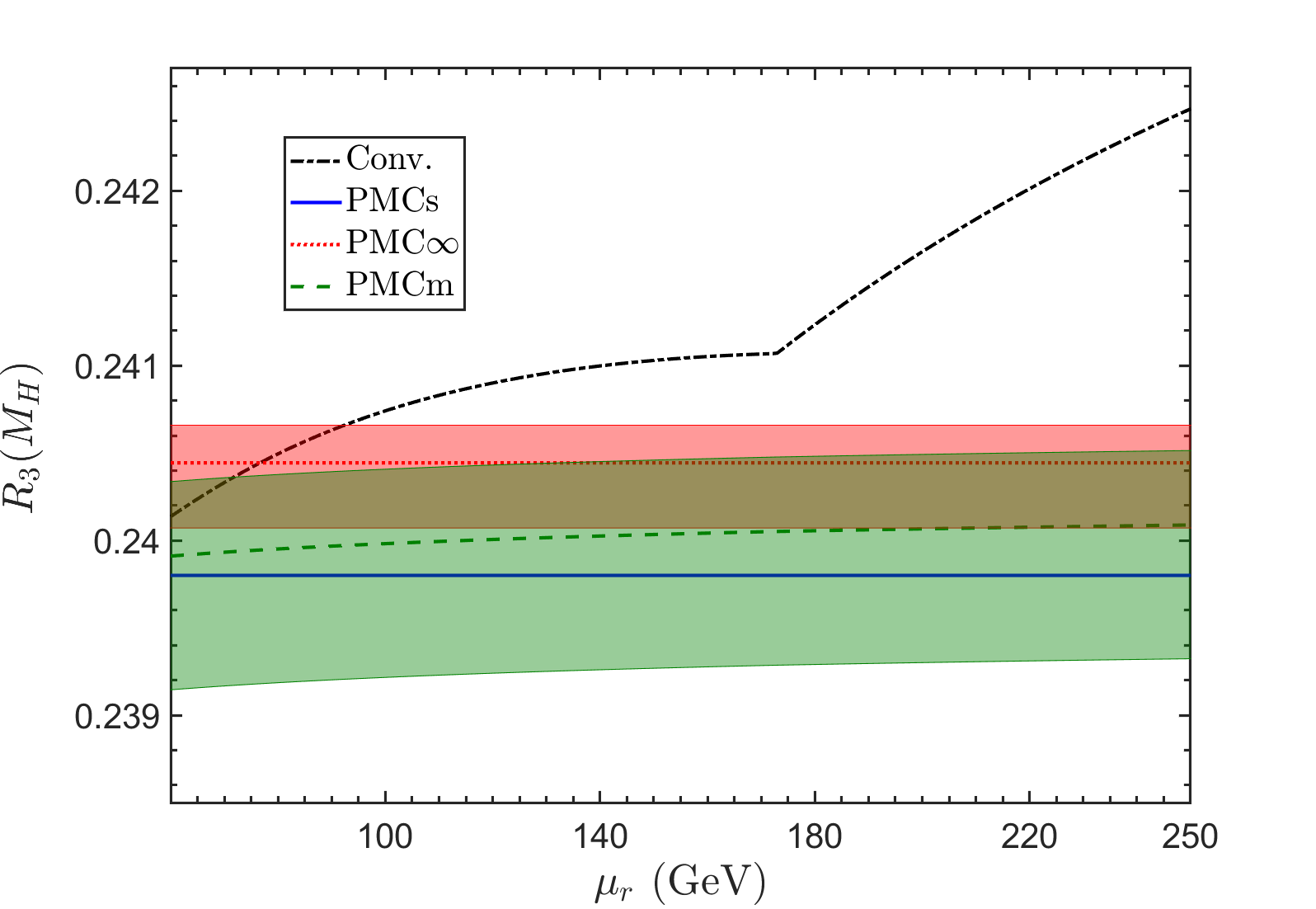}
\caption{The renormalization scale dependence of the four-loop prediction $\tilde{R}_{3}(M_H)$ using the conventional, PMCm, PMCs and PMC$_\infty$ scale-setting procedures. The green/lighter band represents the \textit{second kind of residual scale dependence} under the PMCm, which is obtained by changing the undetermined $Q_4$ to be $2 Q_3$, $(1/2)Q_3$, $Q_1$, $Q_2$ and $(Q_1+Q_2+Q_3)/3$, respectively. The red/darker band represents the \textit{second kind of residual scale dependence} under the PMC$_\infty$, which is obtained by varying the undetermined PMC$_\infty$ scale $\mu_{\rm IV}$ within the region of $[1/2M_H, 2M_H]$.} \label{urdepen3}
\end{figure}

We present the renormalization scale ($\mu_r$) dependence of the four-loop predictions $R_{3}(Q=31.6~{\rm GeV})$, $\hat{R}_{3}(M_{\tau})$, and $\tilde{R}_{3}(M_H)$ using the conventional, PMCm, PMCs, and PMC$_\infty$ (or equivalently, PMCa) scale-setting procedures in Figs.~\ref{urdepen1}, \ref{urdepen2}, and \ref{urdepen3}, respectively. In these figures, we show the \textit{second kind of residual scale dependence} of $R_{3}(Q=31.6~{\rm GeV})$, $\hat{R}_{3}(M_{\tau})$ under the PMCm and PMC$_\infty$ scale-setting procedures with the shaded bands. The green/lighter bands are obtained by changing the undetermined $Q_4$ to $2 Q_3$, $(1/2)Q_3$, $Q_1$, $Q_2$ and $(Q_1+Q_2+Q_3)/3$. And the red/darker bands are obtained by varying the undetermined PMC$_\infty$ scale $\mu_{\rm IV}$ within the region of $[1/2Q, 2Q]$ for $R_{3}(Q=31.6~{\rm GeV})$, $[1{\rm GeV}, 2M_{\tau}]$ for $\hat{R}_{3}(M_{\tau})$, and $[1/2M_H, 2M_H]$ for $\tilde{R}_{3}(M_H)$. Fig.~\ref{urdepen1} shows that the theoretical predictions are smaller than the experimental result. This is reasonable since we have adopted the world average $\alpha_s(M_Z)=0.1179$~\cite{Workman:2022zbs} to set $\Lambda_{\rm QCD}$ for all those observables, and if we adopt the strong coupling $\alpha_s(M_Z)$ fixed by using the data of $e^+e^-$ annihilation alone, we shall obtain consistent predictions in agreement with the data. For example, if using $\alpha_s(M_Z)=0.1224$~\cite{Dissertori:2009ik} that is fixed by using the hadronic event shapes in $e^+e^-$ annihilation to set $\Lambda_{\rm QCD}$, we shall obtain a larger $R_{3}(Q=31.6~{\rm GeV})$, e.g. $R_{3}(Q=31.6~{\rm GeV})=0.04826$ for the PMCs approach, which is consistent with $R^{\rm Exp.}(Q=31.6~{\rm GeV})=0.0527\pm 0.0050$ with errors. It has been noticed that the \textit{second kind of residual scale dependence} of $R_{3}(Q=31.6~{\rm GeV})$ under the PMCm and PMC$_\infty$ scale-setting procedure are both very small, since the $\alpha_s^4$ order correction is highly suppressed in $R_{3}(Q=31.6~{\rm GeV})$. These figures show that by including enough higher-order terms,
\begin{itemize}
\item The renormalization scale dependence of conventional prediction depends strongly on the convergence of perturbative series and the cancellation of scale dependence among different orders. For a numerically strong convergent series such as $R_{3}(Q=31.6~{\rm GeV})$ and $\tilde{R}_{3}(M_H)$, the net scale dependence is only parts per thousand for a wide range of scale choices. For a less convergent series such as $\hat{R}_{3}(M_{\tau})$, the net renormalization scale uncertainty is sizable, which is up to $\sim 18\%$ for $\mu_r\in [1{\rm GeV}, 2M_{\tau}]$, $\sim 24\%$ for $\mu_r\in [1{\rm GeV}, 3M_{\tau}]$, and $\sim 28\%$ for $\mu_r\in [1{\rm GeV}, 5M_{\tau}]$;

\item The PMCm predictions have two kinds of residual scale dependence due to unknown terms. The second residual scale dependence can be greatly suppressed by the extra requirement of conformal invariance; the first property then dominates the net residual scale dependence. For numerically convergent series such as $R_{3}(Q=31.6~{\rm GeV})$ and $\tilde{R}_{3}(M_H)$, the residual scale dependence are small which are less than four parts per thousand~\footnote{ As a comparison, the conventional scale dependence of $\tilde{R}_{3}(M_H)$ is about nine parts per thousand for $\mu_r\in[M_H/2, 2M_H]$}. For a less convergent series such as $\hat{R}_{3}(M_{\tau})$, due to the large residual scale dependence of the NLO-terms, its net residual scale dependence is sizable, which is $\sim 23\%$ for $\mu_r\in [1{\rm GeV}, 2M_{\tau}]$, $\sim 28\%$ for $\mu_r\in [1{\rm GeV}, 3M_{\tau}]$, and $\sim 32\%$ for $\mu_r\in [1{\rm GeV}, 5M_{\tau}]$. Even though in some special cases such as $\hat{R}_{3}(M_{\tau})$, the residual scale dependence may be comparable to the conventional prediction, the PMCm series has no renormalon divergence and it generally has a better pQCD convergence. For the case of $R_{3}(Q=31.6~{\rm GeV})$ and $\tilde{R}_{3}(M_H)$, the PMCm predictions show weaker dependence on $\mu_r$ and its prediction can be more accurate than conventional pQCD predictions;

\item The PMC$_\infty$ and PMCa predictions only have the \textit{second kind of residual scale dependence}, which are suppressed for the present four-loop predictions. The magnitude of the residual scale dependence depends on the convergence of the resultant series, and for the present processes, the \textit{second kind of residual scale dependence} are only about parts per thousand to a few percent. Due to the application of ``intrinsic conformality" or equivalently the requirement of scale invariance at each order, the determined PMC$_\infty$ and PMCa scales are not of the perturbative nature, but they can be very small in certain cases. For the case of $R_{3}(Q=31.6~{\rm GeV})$, we obtain a much small scale $\mu_{\rm III}=0.003$ GeV, which is unreasonable and indicates the PMC$_\infty$ approach may not be applicable for this process. To do a numerical estimation, we have adopted the MPT model to calculate the magnitude of $\alpha_s$ at such small scale; Fig.~\ref{urdepen1} shows the MPT prediction deviates from other approaches by about $15\%$; By including the uncertainty from the MPT model parameter $\xi=10\pm2$, the ${\rm PMC}_{\infty}$ prediction is still deviates from other approaches by about $11\%$;

\item The PMCs predictions for the dependence of observables on the renormalization scale are flat lines. The \textit{first kind of residual scale dependence} of the PMCs predictions only affects the precision of the magnitude of effective $\alpha_s$, and the PMCs predictions are exactly independent to the choice of $\mu_r$ at any fixed order.
\end{itemize}

\section{Summary}

The PMC provides a rigorous first principle method to eliminate conventional renormalization scheme and scale ambiguities for high-momentum transfer processes. Its predictions have a solid theoretical foundation, satisfying renormalization group invariance and all other self-consistency conditions derived from the renormalization group. The PMC has now been successfully applied to many high-energy processes.

The original PMCm approach is a multi-scale-setting approach, which introduces individual scales at each order which are determined by using the non-conformal $\{\beta_i\}$-terms recursively. It determines the correct magnitudes of the effective $\alpha_s$ at each order; the corresponding PMC scales thus reflect the varying virtuality of the amplitudes at each order. Due to the unknown higher-order terms, the PMCm approach has two kinds of residual scale dependence. The residual scale dependence of the PMCm predictions are generally small, and retain both the $\alpha_s$-power suppression and the exponential suppression. However for low-order pQCD predictions, they could be sizable due to a possibly weaker pQCD convergence for the perturbative series of the PMC scale or the pQCD approximant. Several alternative PMC approaches, such as PMCs, PMC$_\infty$ and PMCa, have been proposed to further suppress the residual scale dependence. In this paper, we have given a detailed comparison of four PMC-like scale-setting procedures for three quantities $R_{e^+ e^-}$, $R_{\tau}$, and $\Gamma(H\to b\bar{b})$ up to four-loop pQCD order. All of these PMC-like approaches have the same conformal coefficients, but because the different way of eliminating the $\{\beta_i\}$-terms and large log-terms vary, the effective scales are different for the different approaches. Moreover, we observe that
\begin{itemize}
\item[$\circ$] The PMCs approach determines an overall effective $\alpha_s$ by eliminating all the RG-dependent nonconformal $\{\beta_i\}$-terms; this results in a single effective scale which effectively replaces the individual PMC scales of PMCm approach in the sense of a mean value theorem. The PMCs prediction is renormalization scale-and-scheme independent up to any fixed order. The \textit{first kind of residual scale dependence} is highly suppressed, since the PMC scale at all known orders is determined at the same highest order accuracy. There is no \textit{second kind of residual scale dependence}. The PMCs prediction also avoids the small scale problem which sometimes emerges in the multi-scale approaches;

\item[$\circ$] The PMC$_\infty$ approach fixes the PMC scales at each order by using the property of intrinsic conformality, which ensures the scale invariance of the pQCD series at each order. The resulting PMC scales have no ambiguities, are not of the perturbative nature, and thus avoid the \textit{first kind of residual scale dependence}. Since the last effective scale of the highest order perturbative term is not determined, the PMC$_\infty$ prediction still has the \textit{second kind of residual scale dependence}. When more loop terms have been included, and there is no very small scale problem, it will give similar predictions as those of PMCm and PMCs;

\item[$\circ$] The PMCa approach fixes the PMC scales at each order by requiring all the scale-dependent terms at each order to vanish in a step-by-step way. We have demonstrated that PMCa and PMC$_\infty$ are equivalent to each other. This equivalence reflects the fact that intrinsic conformality requires the scale invariance of the pQCD series at each perturbative order, and vice versa.
\end{itemize}

The PMCs approach is close to PMCm approach in achieving the goals of the PMC by inheriting most of the features of the PMCm approach: It uses the RGE recursively by using a weighted average of $\{\beta_i\}$-terms at different orders; its predictions are again exactly scale independent, and the convergence of the pQCD expansion is greatly improved due to the elimination of the divergent renormalon terms. The resulting relations between physical observables are also independent of the choice of the renormalization scheme. Any residual scale dependence of PMCm is highly suppressed. Due to the resulting conformal series and the highly precise determination of the magnitude of the effective $\alpha_s$, it provides a reliable basis for predicting unknown higher-order contributions and thus achieves precise pQCD predictions for observables such as $R_{e^+e^-}$, $R_{\tau}$ and $\Gamma(H \to b \bar{b})$ without renormalization scale ambiguities.

\hspace{2cm}

\noindent {\bf Acknowledgments:} We thank Sheng-Quan Wang for helpful discussions. This work was supported in part by the Natural Science Foundation of China under Grant No.11905056, No.12047506, 12147102, No.12175025 and No.12247129, the Fundamental Research Funds for the Central Universities under Chongqing Graduate Research and Innovation under No.ydstd1912, and the Department of Energy contract DE--AC02--76SF00515. SLAC-PUB-17623.

\end{document}